\documentclass[journal,onecolumn,10pt,twoside]{IEEEtranTCOM}
\newtheorem{definition}{{Definition}}
\newtheorem{theorem}{{Theorem}}
\newtheorem{lemma}{{Lemma}}

\newtheorem{corollary}{{Corollary}}

\usepackage{cite}
\usepackage{amsmath,amssymb,amsfonts}
\usepackage{algorithmic}
\usepackage{graphicx}
\usepackage{textcomp}
\usepackage{xcolor}

\normalsize

\ifCLASSINFOpdf
\else
\fi

\begin{document}
%
\title{Gaussian Rate-Distortion-Perception Coding and Entropy-Constrained Scalar Quantization}
%
%
%

\author{Li~Xie, Liangyan~Li,  Jun~Chen, Lei~Yu,
        and~Zhongshan~Zhang}
\maketitle

\begin{abstract}
This paper investigates the best known bounds on the quadratic Gaussian  distortion-rate-perception function with limited common randomness for the Kullback-Leibler divergence-based perception measure, as well as  their counterparts for the squared Wasserstein-2 distance-based perception measure, recently established by Xie et al. These bounds are shown to be  nondegenerate in the sense that they cannot be deduced from each other via a refined version of Talagrand's transportation inequality. On the other hand, an improved lower bound is established when the perception measure is given by the squared Wasserstein-2 distance. In addition, 
it is revealed by exploiting the connection between rate-distortion-perception coding and entropy-constrained scalar quantization that all the aforementioned bounds are generally not tight in the weak perception constraint regime.
\end{abstract}


%
\IEEEpeerreviewmaketitle

\section{Introduction}

\IEEEPARstart{R}{ate}-distortion-perception theory \cite{Matsumoto18, Matsumoto19, BM19,TW21, YWYML21, ZQCK21, QZCK22, LZCK22, LZCK22_2, CYWSGT22, SPCYK23,QSCKYSGT24,SCKY24}, as a generalization of Shannon's rate-distortion theory, has received considerable attention in recent years. It provides a framework for investigating the performance limits of perception-aware image compression. This is partly accomplished by assessing compression results more comprehensively, using both distortion and perception measures. Unlike distortion measures, which compare each compressed image with its corresponding source image, perception measures focus on the ensemble-level relationship between pre- and post-compression images. It has been observed that at a given coding rate, there exists a tension between distortion loss and  perception loss \cite{BM18,FMM21,FWM24}. Moreover,  the presence of a perception constraint often necessitates the use of stochastic algorithms \cite{TA21,Wagner22,HWG24}. In contrast, deterministic algorithms are known to be adequate for conventional lossy source coding.

Although significant progress has been made in characterizing the information-theoretic limits of rate-distortion-perception coding, existing results are almost exclusively restricted to  special scenarios with the availability of unlimited common randomness or with the perfect perception constraint (also referred to as perfect realism). 
To the best of our knowledge, the only exception is \cite{XLCZ24}, which makes an initial attempt to study the fundamental distortion-rate-perception tradeoff 
with limited common randomness by leveraging the research findings from  output-constrained lossy source coding \cite{LKK10,LKK11,KZLK13,SLY15J1,SLY15J2}. In particular, lower and upper bounds on the quadratic Gaussian distortion-rate-perception function under a specified amount of common randomness
 are established in \cite{XLCZ24} for both  Kullback-Leibler divergence-based  and squared Wasserstein-2 distance-based perception measures.
 These bounds shed light on the utility of common randomness as a resource in  rate-distortion-perception coding, especially when the perceptual quality is not required to be perfect.  
 On the other hand, they in general do not match and  are  therefore inconclusive. Note that the aforementioned upper bounds are derived by restricting the reconstruction distribution to be Gaussian. A natural question thus arises whether this restriction incurs any penalty. A negative answer to this question is equivalent to the existence of some new Gaussian extremal inequalities, which are of independent interest. It is also worth noting that Kullback-Leibler divergence and squared Wasserstein-2 distance are related via Talagrand's transporation inequality \cite{Talagrand96} when the reference distribution is Gaussian. As such, there exists an intrinsic connection between the quadratic distortion-rate-perception functions associated with Kullback-Leibler divergence-based  and squared Wasserstein-2 distance-based perception measures.  
 This connection has not been explored in existing literature.

 We shall show that the bounds on the quadratic Gaussian distortion-rate-perception function with limited common randomness for the  Kullback-Leibler divergence-based perception measure cannot be deduced from their counterparts for the  squared Wasserstein-2 distance-based perception measure  via a refined version of Talagrand's transportation inequality. In this sense, they are not degenerate. On the other hand, it turns out that the lower bound can be improved via an additional tunable parameter when the perception measure is given by the squared Wasserstein-2 distance. Furthermore,  all the aforementioned bounds are  generally not tight in the weak perception constraint regime. 
 We demonstrate this result by exploiting the connection between rate-distortion-perception coding and entropy-constrained scalar quantization. Our finding implies 
  that restricting the reconstruction distribuiton to be Gaussian may incur a penalty.  This is somewhat surprising in view of the fact that the quadratic Gaussian distortion-rate-percetion function with limited common randomness admits a single-letter characterization, which often implies the existence of a corresponding Gaussian extremal inequality \cite{GN14}.

 The rest of this paper is organized as follows. Section \ref{sec:Definition} contains the  definition of quadratic distortion-rate-perception function with limited common randomness and a review of some relevant results.  Our technical contributions are presented in Sections \ref{sec:Talagrand}, \ref{sec:lowerbound}, and \ref{sec:ECSQ}. We conclude the paper in Section \ref{sec:conclusion}.

We adopt the standard notation for information measures, e.g.,  $H(\cdot)$ for entropy, $h(\cdot)$ for differential entropy,  $I(\cdot;\cdot)$ for mutual information, and $J(\cdot)$ for Fisher information.
The cardinality of set $\mathcal{S}$ is denoted by $|\mathcal{S}|$. For a given random variable $X$, its distribution, mean, and variance are written as $p_X$, $\mu_X$, and $\sigma^2_X$, respectively. We use  $\Pi(p_X,p_{\hat{X}})$ to represent  the set of all possible couplings of $p_X$ and $p_{\hat{X}}$.
For real numbers $a$ and $b$, let $a\wedge b:=\min\{a,b\}$, $a\vee b:=\max\{a,b\}$, and $(a)_+:=\max\{a,0\}$.
 Throughout this paper, the logarithm function is
assumed to have base $e$.





\section{Problem Definition and Existing Results}\label{sec:Definition}

\begin{figure}[htbp]
	\centerline{\includegraphics[width=12cm]{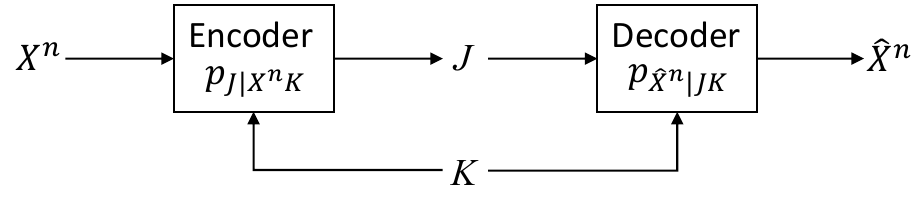}} \caption{System diagram.}
	\label{fig:system} 
\end{figure}

A length-$n$ rate-distortion-perception coding system (see Fig. \ref{fig:system}) consists of an encoder $f^{(n)}:\mathbb{R}^n\times\mathcal{K}\rightarrow\mathcal{J}$, a  decoder $g^{(n)}:\mathcal{J}\times\mathcal{K}\rightarrow\mathbb{R}^n$, and a random seed $K$. It takes an i.i.d. source sequence $X^n$ as input and produces an i.i.d. reconstruction sequence $\hat{X}^n$. Specifically, the encoder
 maps   $X^n$ and  $K$ to a codeword $J$ in codebook $\mathcal{J}$ according to some conditional distribution $p_{J|X^nK}$ while the decoder  generates  $\hat{X}^n$ based on $J$ and $K$ according to some conditional distribution $p_{\hat{X}^n|JK}$. Here, $K$  is assumed to be uniformly distributed over the alphabet $\mathcal{K}$ and independent of $X^n$. The end-to-end distortion is quantified by $\frac{1}{n}\sum_{t=1}^n\mathbb{E}[(X_t-\hat{X}_t)^2]$ and the perceptual quality by $\frac{1}{n}\sum_{t=1}^n\phi(p_{X_t},p_{\hat{X}_t})$ with some divergence $\phi$. It is clear that $\frac{1}{n}\sum_{t=1}^n\phi(p_{X_t},p_{\hat{X}_t})=\phi(p_X,p_{\hat{X}})$, where $p_X$ and $p_{\hat{X}}$ are the marginal distributions of $X^n$ and $\hat{X}^n$, respectively.


\begin{definition}
	For an i.i.d. source $\{X_t\}_{t=1}^{\infty}$, distortion level $D$ is said to be achievable   subject to the compression rate constraint $R$, the common randomness rate constraint $R_c$, and the perception constraint $P$	
	if there exists a length-$n$ rate-distortion-perception coding system such that 	
	\begin{align}
		&\frac{1}{n}\log|\mathcal{J}|\leq R,\\
		&\frac{1}{n}\log|\mathcal{K}|\leq R_c,\\
		&\frac{1}{n}\sum\limits_{t=1}^n\mathbb{E}[(X_t-\hat{X}_t)^2]\leq D,\\
		&\frac{1}{n}\sum\limits_{t=1}^n\phi(p_{X_t},p_{\hat{X}_t})\leq P,
	\end{align}
	and the reconstruction sequence $\hat{X}^n$ is ensured to be i.i.d. The infimum of such achievable distortion levels $D$ is denoted by  $D(R,R_c,P|\phi)$.
\end{definition}

The following result \cite[Theorem 1]{XLCZ24}, which is built upon  \cite[Theorem 1]{SLY15J2} (see also \cite[Theorem 2]{Wagner22}),  provides a single-letter characterization of $D(R,R_c,P|\phi)$.

\begin{theorem}\label{thm:outputconstrained}
	For $p_{X}$ with $\mathbb{E}[X^2]<\infty$,
	\begin{align}
		D(R,R_c,P|\phi)&=\inf\limits_{p_{U\hat{X}|X}}\mathbb{E}[(X-\hat{X})^2]\label{eq:inf}\\
	\mbox{subject to }	&\quad X\leftrightarrow U\leftrightarrow\hat{X}\mbox{ form a Markov chain},\label{eq:const1}\\ 
		&\quad I(X;U)\leq R,\label{eq:const2}\\
		&\quad I(\hat{X};U)\leq R+R_c,\label{eq:const3}\\
		&\quad \phi(p_X,p_{\hat{X}})\leq P.\label{eq:const4}
	\end{align}
\end{theorem}

Explicit lower and upper bounds on $D(R,R_c,P|\phi)$ are established for $p_X=\mathcal{N}(\mu_X,\sigma^2_X)$ when $\phi(p_X,p_{\hat{X}})=\phi_{KL}(p_{\hat{X}}\|p_X)$ \cite[Theorem 3]{XLCZ24} or $\phi(p_X,p_{\hat{X}})=W^2_2(p_X,p_{\hat{X}})$ \cite[Theorem 4]{XLCZ24},
where
\begin{align}
	\phi(p_{\hat{X}}\|p_X):=\mathbb{E}\left[\log\frac{p_{\hat{X}}(\hat{X})}{p_X(\hat{X})}\right]
\end{align}
is the Kullback-Leibler divergence and 
\begin{align}
	W^2_2(p_X,p_{\hat{X}}):=\inf\limits_{p_{X\hat{X}}\in\Pi(p_X,p_{\hat{X}})}\mathbb{E}[(X-\hat{X})^2]
\end{align}
is the squared  Wasserstein-2 distance. Let 
\begin{align}
	\xi(R,R_c):=\sqrt{(1-e^{-2R})(1-e^{-2(R+R_c)})}.
\end{align}
Moreover, let
\begin{align}
	\psi(\sigma_{\hat{X}}):=\log\frac{\sigma_X}{\sigma_{\hat{X}}}+\frac{\sigma^2_{\hat{X}}-\sigma^2_X}{2\sigma^2_X}\label{eq:defpsi}
\end{align}
and
$\sigma(P)$ be the unique number  $\sigma\in[0,\sigma_X]$ satisfying $\psi(\sigma)=P$.



\begin{theorem}\label{thm:KL}
	For  $p_X=\mathcal{N}(\mu_X,\sigma^2_X)$,
	\begin{align}
		\underline{D}(R,R_c,P|\phi_{KL})\leq D(R,R_c,P|\phi_{KL})\leq\overline{D}(R,R_c,P|\phi_{KL}),
	\end{align}
where
\begin{align}
		&\underline{D}(R,R_c,P|\phi_{KL}):=\min\limits_{\sigma_{\hat{X}}\in[\sigma(P),\sigma_X]}\sigma^2_X+\sigma^2_{\hat{X}}-2\sigma_X\sigma_{\hat{X}}\sqrt{(1-e^{-2R})(1-e^{-2(R+R_c+P-\psi(\sigma_{\hat{X}}))})}
	\end{align}
and
\begin{align}
	&\overline{D}(R,R_c,P|\phi_{KL}):=\sigma^2_X-\sigma^2_X\xi^2(R,R_c)+(\sigma(P)-\sigma_X\xi(R,R_c))^2_+.
	\end{align}
\end{theorem}

\begin{theorem}\label{thm:W2}
For  $p_X=\mathcal{N}(\mu_X,\sigma^2_X)$,
\begin{align}
	\underline{D}(R,R_c,P|W^2_2)\leq D(R,R_c,P|W^2_2)\leq\overline{D}(R,R_c,P|W^2_2),
\end{align}
where
\begin{align}
	&\underline{D}(R,R_c,P|W^2_2):=\min\limits_{\sigma_{\hat{X}}\in[(\sigma_X-\sqrt{P})_+,\sigma_X]}\sigma^2_X+\sigma^2_{\hat{X}}-2\sigma_X\sqrt{(1-e^{-2R})(\sigma^2_{\hat{X}}-(\sigma_Xe^{-(R+R_c)}-\sqrt{P})^2_+)}
\end{align}	
and
\begin{align}
	&\overline{D}(R,R_c,P|W^2_2):=\sigma^2_X-\sigma^2_X\xi^2(R,R_c)+(\sigma_X-\sqrt{P}-\sigma_X\xi(R,R_c))^2_+.
\end{align}
\end{theorem}

The next three sections are devoted to investigating the tightness of these bounds, which will shed light on 
rate-distortion-perception coding in general.

\section{Kullback-Leibler Divergence vs. Squared Wasserstein-2 Distance}\label{sec:Talagrand}

For $p_X=\mathcal{N}(\mu_X,\sigma^2_X)$, Talagrand's transportation inequality \cite{Talagrand96} states that
\begin{align}
	W^2_2(p_X,p_{\hat{X}})\leq 2\sigma^2_X\phi_{KL}(p_{\hat{X}}\|p_X),\label{eq:originalTalagrand}
\end{align}
which immediately implies
\begin{align}
	D(R,R_c,2\sigma^2_XP|W^2_2)\leq D(R,R_c,P|\phi_{KL}).\label{eq:weakconnection}
\end{align}
Note that Talagrand's transportation inequality does not impose any assumptions on $p_{\hat{X}}$. However, when $p_X=\mathcal{N}(\mu_X,\sigma^2_X)$, it suffices to consider $p_{\hat{X}}$ with $\mu_{\hat{X}}=\mu_X$ and $\sigma_{\hat{X}}\leq\sigma_X$ as far as $D(R,R_c,P|\phi_{KL})$ and $D(R,R_c,P|W^2_2)$ are concerned \cite[Lemmas 1 and 3]{XLCZ24}. With this restriction on $p_{\hat{X}}$, we have the following refined version of Talagrand's transportation inequality, which leads to an improvement on (\ref{eq:weakconnection}).

\begin{theorem}\label{thm:KLW2} 
For $p_X=\mathcal{N}(\mu_X,\sigma^2_X)$ and $p_{\hat{X}}$ with $\mu_{\hat{X}}=\mu_X$ and $\sigma_{\hat{X}}\leq\sigma_X$,
\begin{align}
W^2_2(p_X,p_{\hat{X}})\leq 2\sigma^2_X(1-e^{-\phi_{KL}(p_{\hat{X}}\|p_X)}).\label{eq:refinedTalagrand}
\end{align}
As a consequence, 
	\begin{align}
	D(R,R_c,2\sigma^2_X(1-e^{-P})|W^2_2)\leq  D(R,R_c,P|\phi_{KL})\label{eq:strongconnection}
\end{align}
when $p_X=\mathcal{N}(\mu_X,\sigma^2_X)$.
\end{theorem}
\begin{IEEEproof}
	See Appendix \ref{app:KLW2}.
\end{IEEEproof}


It is clear that (\ref{eq:refinedTalagrand}) and (\ref{eq:strongconnection}) are  stronger than their counterparts in  (\ref{eq:originalTalagrand}) and (\ref{eq:weakconnection}) since $1+z\leq e^z$ for all $z$. Theorem \ref{thm:KLW2} implies that for $p_X=\mathcal{N}(\mu_X,\sigma^2_X)$, 	
	every lower bound on $D(R,R_c,\cdot|W^2_2)$ induces a lower bound on $D(R,R_c,\cdot|\phi_{KL})$ and every upper bound on $D(R,R_c,\cdot|\phi_{KL})$ induces an upper bound on $D(R,R_c,\cdot|W^2_2)$; in particular,  we have
	\begin{align}
		D(R,R_c,P|\phi_{KL})\geq\underline{D}(R,R_c,2\sigma^2_X(1-e^{-P})|W^2_2)\label{eq:induced1}
		\end{align}
	and
	\begin{align}
		D(R,R_c,P|W^2_2)\leq\overline{D}(R,R_c,\nu(P)|\phi_{KL}),\label{eq:induced2}
	\end{align}
where
\begin{align}
	\nu(P)&:=\log\frac{2\sigma^2_X}{(2\sigma^2_X-P)_+}.
\end{align}
It is thus of considerable interest to see how these induced bounds are compared to their counterparts in Theorems \ref{thm:KL} and \ref{thm:W2}, namely, 
	\begin{align}
		D(R,R_c,P|\phi_{KL})\geq\underline{D}(R,R_c,P|\phi_{KL})\label{eq:direct1}
		\end{align}
	and
	\begin{align}
		D(R,R_c,P|W^2_2)\leq\overline{D}(R,R_c,P|W^2_2).\label{eq:direct2}
	\end{align}
	The following result indicates that (\ref{eq:induced1}) and (\ref{eq:induced2}) are in general looser. In this sense, (\ref{eq:direct1}) and (\ref{eq:direct2}) are nondegenerate.

\begin{theorem}\label{thm:KLvsW2}
	For $p_X=\mathcal{N}(\mu_X,\sigma^2_X)$,
	\begin{align}
			\underline{D}(R,R_c,P|\phi_{KL})\geq\underline{D}(R,R_c,2\sigma^2_X(1-e^{-P})|W^2_2)\label{eq:W2toKL}
			\end{align}
		and
		\begin{align}
		\overline{D}(R,R_c,P|W^2_2)\leq\overline{D}(R,R_c,\nu(P)|\phi_{KL}).\label{KLtoW2}
	\end{align}
\end{theorem}
\begin{IEEEproof}
	See Appendix \ref{app:nondegenerate}.		
\end{IEEEproof}

\begin{figure}[htbp]
		\centerline{\includegraphics[width=12cm]{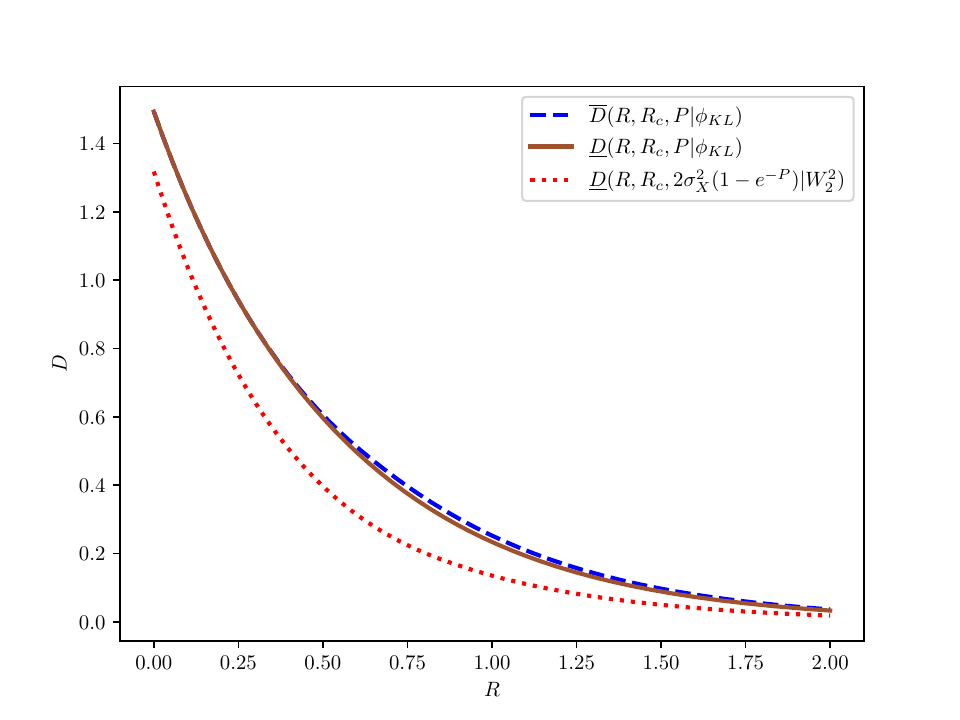}} \caption{Illustrations of $\overline{D}(R,R_c,P|\phi_{KL})$, $\underline{D}(R,R_c,P|\phi_{KL})$, and $\underline{D}(R,R_c,2\sigma^2_X(1-e^{-P})|W^2_2)$ for  $p_X=\mathcal{N}(0,1)$, $R_c=0$, and $P=0.1$.}
		\label{fig:correspondence_lower} 
	\end{figure}

\begin{figure}[htbp]
	\centerline{\includegraphics[width=12cm]{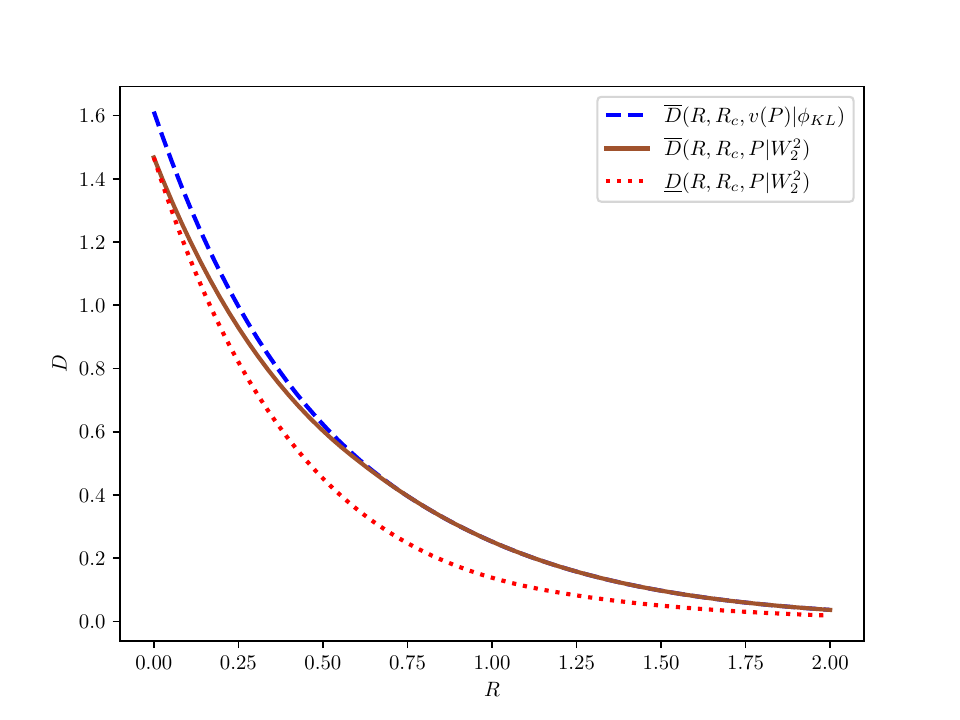}} \caption{Illustrations of $\overline{D}(R,R_c,\nu(P)|\phi_{KL})$, $\overline{D}(R,R_c,P|W^2_2)$, and $\underline{D}(R,R_c,P|W^2_2)$  for  $p_X=\mathcal{N}(0,1)$, $R_c=0$, and $P=0.1$.}
	\label{fig:correspondence_upper} 
\end{figure}

	It be can seen from Fig. \ref{fig:correspondence_lower} that $\underline{D}(R,R_c,2\sigma^2_X(1-e^{-P})|W^2_2)$ is indeed a looser lower bound on $D(R,R_c,P|\phi_{KL})$ as compared to 
	$\underline{D}(R,R_c,P|\phi_{KL})$ and the latter almost meets the upper bound $\overline{D}(R,R_c,P|\phi_{KL})$. 	
	Similarly, Fig. \ref{fig:correspondence_upper} shows that  $\overline{D}(R,R_c,\nu(P)|\phi_{KL})$ is indeed a  looser upper bound on $D(R,R_c,P|W^2_2)$ 
	 as compared to $\overline{D}(R,R_c,P|W^2_2)$, especially in the low rate regime, where the latter has a diminishing gap from the lower bound $\underline{D}(R,R_c,P|W^2_2)$.
	 
	
\section{An Improved Lower Bound}\label{sec:lowerbound}

The main result of this section is the following improved lower bound on $D(R,R_c,P|W^2_2)$.

\begin{theorem}\label{thm:improvedlowerbound}
	For $p_X=\mathcal{N}(\mu_X,\sigma^2_X)$,
	\begin{align}
		D(R,R_c,P|W^2_2)\geq\underline{D}'(R,R_c,P|W^2_2)\geq\underline{D}(R,R_c,P|W^2_2),\label{eq:twoinequalities}
	\end{align}
where	
	\begin{align}
		&\underline{D}'(R,R_c,P|W^2_2):=\min\limits_{\sigma_{\hat{X}}\in[(\sigma_X-\sqrt{P})_+,\sigma_X]}\sup\limits_{\alpha>0} \sigma^2_X+\sigma^2_{\hat{X}}-2\sigma_X\sqrt{(1-e^{-2R})(\sigma^2_{\hat{X}}-\delta^2_+(\sigma_{\hat{X}},\alpha))}\label{eq:minsup}
	\end{align}
with
	\begin{align}
	&\delta_+(\sigma_{\hat{X}},\alpha):=\frac{(\sigma_Xe^{-(R+R_c)}-\sqrt{\sigma^2_X-\alpha(\sigma^2_X+\sigma^2_{\hat{X}}-P)+\alpha^2\sigma^2_{\hat{X}}})_+}{\alpha}.
	\end{align}
Moreover, the second inequality in (\ref{eq:twoinequalities}) is strict if and only if $R\in(0,\infty)$, $R_c\in(0,\infty)$, and $P\in(0,\sigma^2_X(2-e^{-2R}-2\sqrt{(1-e^{-2R})(1-e^{-2(R+R_c)})}))$.
\end{theorem}
\begin{IEEEproof}
See Appendix \ref{app:improvedlowerbound}.
\end{IEEEproof}




\begin{figure}[htbp]
	\centerline{\includegraphics[width=12cm]{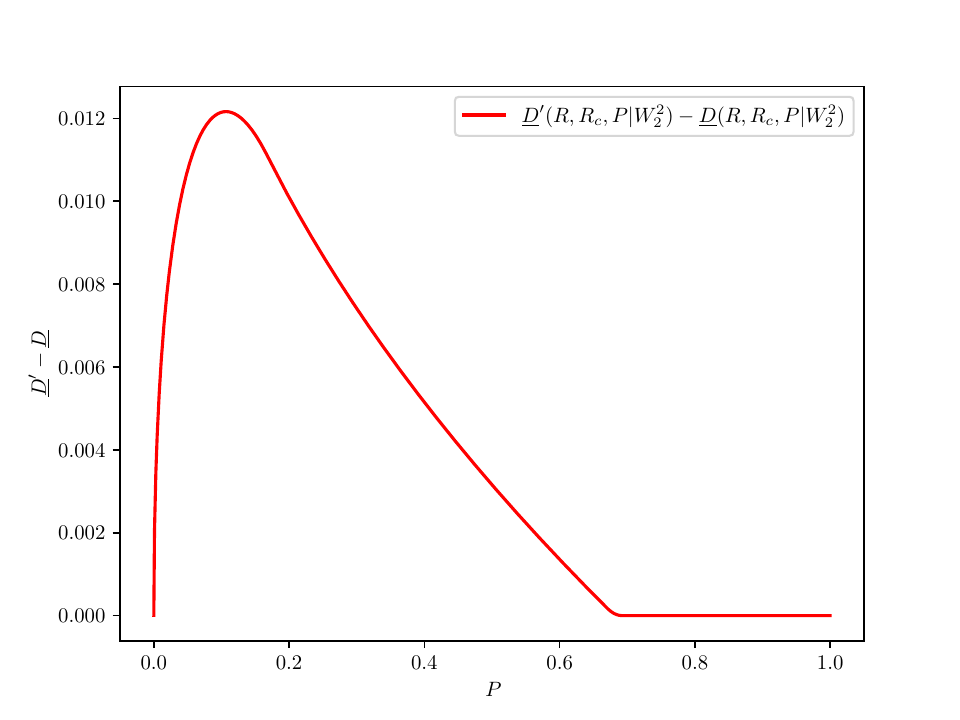}} \caption{Illustrations of $\underline{D}'(R,R_c,P|W^2_2)-\underline{D}(R,R_c,P|W^2_2)$ for  $p_X=\mathcal{N}(0,1)$, $R=0.1$,  and $R_c=0.1$.}\label{fig:diffP}
\end{figure}

\begin{figure}[htbp]
	\centerline{\includegraphics[width=12cm]{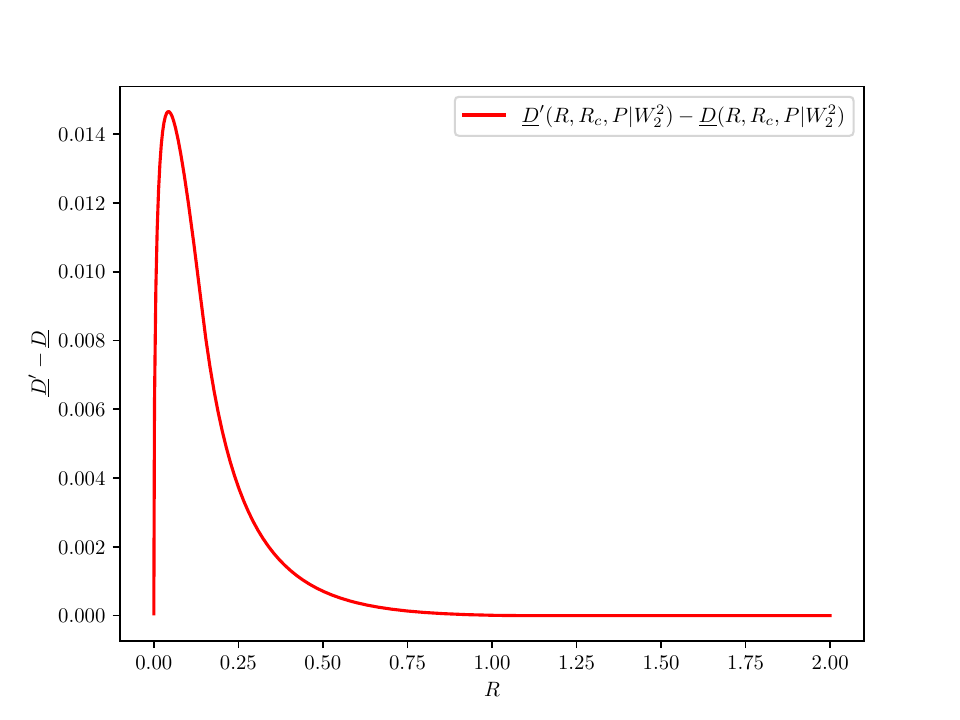}} \caption{Illustrations of  $\underline{D}'(R,R_c,P|W^2_2)-\underline{D}(R,R_c,P|W^2_2)$ for  $p_X=\mathcal{N}(0,1)$, $R_c=0.1$, and $P=0.1$.}\label{fig:diffR}
\end{figure}


The difference between $\underline{D}'(R,R_c,P|W^2_2)$ and $\underline{D}(R,R_c,P|W^2_2)$ against $P$ is plotted in Fig. \ref{fig:diffP} for the case  $p_X=\mathcal{N}(0,1)$, $R=0.1$,  and $R_c=0.1$.
According to Theorem \ref{thm:improvedlowerbound}, for $R\in(0,\infty)$ and $R_c\in(0,\infty)$, we have $\underline{D}'(R,R_c,P|W^2_2)=\underline{D}(R,R_c,P|W^2_2)$ when 
\begin{align}
	P\geq\sigma^2_X(2-e^{-2R}-2\sqrt{(1-e^{-2R})(1-e^{-2(R+R_c)})}).\label{eq:Pthreshold}
\end{align}
Setting $\sigma^2_X=1$, $R=0.1$, and $R_c=0.1$ in (\ref{eq:Pthreshold}) gives $P\gtrapprox 0.692$, which is consistent with the result shown in Fig. \ref{fig:diffP}. Fig. \ref{fig:diffR} plots the  difference between $\underline{D}'(R,R_c,P|W^2_2)$ and $\underline{D}(R,R_c,P|W^2_2)$ against $R$  for the case  $p_X=\mathcal{N}(0,1)$, $R_c=0.1$, and $P=0.1$. As shown in Appendix \ref{app:Rthreshold}, for $R_c\in(0,\infty)$ and $P\in(0,\infty]$, we can write
(\ref{eq:Pthreshold})  alternatively as
\begin{align}
	R\geq\begin{cases}
	0	&\mbox{if }P\geq\sigma^2_X,\\
	-\frac{1}{2}\log\frac{\zeta_3}{\zeta_2}	&\mbox{if }R_c=\log2, P<\sigma^2_X,\\
	-\frac{1}{2}\log\frac{\zeta_2-\sqrt{\zeta^2_2-4\zeta_1\zeta_3}}{2\zeta_1}	&\mbox{otherwise},
	\end{cases}\label{eq:Rthreshold}
\end{align}
where
\begin{align}
	&\zeta_1:=4e^{-2R_c}-1,\\
	&\zeta_2:=4e^{-2R_c}+\frac{2P}{\sigma^2_X},\\
	&\zeta_3:=\frac{(4\sigma^2_X-P)P}{\sigma^4_X}.
\end{align}
Setting $\sigma^2_X=1$, $R_c=0.1$, and $P=0.1$ in (\ref{eq:Rthreshold}) gives $R\gtrapprox1.052$, which is consistent with the result shown in Fig. \ref{fig:diffR}.


It is interesting to note that for $p_X=\mathcal{N}(\mu_X,\sigma^2_X)$,
	\begin{align*}
	\underline{D}'(R,0,P|W^2_2)&=\underline{D}(R,0,P|W^2_2)\\
	&=\sigma^2_Xe^{-2R}+(\sigma_Xe^{-R}-\sqrt{P})^2_+,
\end{align*}
which coincides with the minimum achievable mean squared error  at rate $R$ and squared Wasserstein-2 perception loss $P$ when the reconstruction sequence is not required to be i.i.d. \cite{YWL22, QCYX24}. As shown in the next section, $\underline{D}(R,0,P|W^2_2)$ and $\underline{D}'(R,0,P|W^2_2)$ are actually strictly below $D(R,0,P|W^2_2)$ for sufficiently large $P$. Therefore, a price has to be paid for enforcing the i.i.d. reconstruction constraint. 
This should be contrasted with the case $R_c=\infty$ for which it is known that the rate-distortion-perception tradeoff remains the same regardless of whether the reconstruction sequence is required to be i.i.d. or not \cite[Theorem 3]{TW21} \cite[Theorem 9]{CYWSGT22}.


\section{Connection with Entropy-Constrained Scalar Quantization}\label{sec:ECSQ}

This section is devoted to  investigating the tightness of bounds in Theorems \ref{thm:KL}, \ref{thm:W2}, and \ref{thm:improvedlowerbound}. We shall focus on the weak perception constraint regime where $P$ is sufficiently large.

To this end, it is necessary to first gain a better understanding of the properties of $D(R,R_c,P|\phi)$. Clearly,  the map $(R,R_c,P)\mapsto D(R,R_c,P|\phi)$ is monotonically decreasing in each of its variables. The following result provides further information regarding $D(R,R_c,P|\phi)$ under certain conditions.
\begin{theorem}\label{thm:property}
	For $p_X$ with bounded support, if $p_{\hat{X}}\mapsto\phi(p_X, p_{\hat{X}})$ is lower semicontinuous in the topology of weak convergence\footnote{It is known that 
			$p_{\hat{X}}\mapsto \phi_{KL}(p_{\hat{X}}\|p_X)$ \cite[Theorem 4.9]{PW24} and $p_{\hat{X}}\mapsto W^2_2(p_X,p_{\hat{X}})$ \cite[Remark 6.12]{Villani08} are  lower semicontinuous in the topology of weak convergence.}, then the infimum in (\ref{eq:inf}) can be attained and the map $(R,R_c,P)\mapsto D(R,R_c,P|\phi)$ is right-continuous in each of its variables.
\end{theorem}
\begin{IEEEproof}
	See Appendix \ref{app:property}.
\end{IEEEproof}

Theorem \ref{thm:property} is not applicable when $p_X$ is a Gaussian distribution. However, it will be seen that assuming the attainability of the infimum in  (\ref{eq:inf}) greatly simplifies the reasoning and helps develop the intuition behind the rigorous proof of the main result in this section (see Theorem \ref{thm:nontightness}).

The next two results deal with the special cases $\phi(p_X,p_{\hat{X}})=\phi_{KL}(p_{\hat{X}}\|p_X)$ and $\phi(p_X,p_{\hat{X}})=W^2_2(p_X,p_{\hat{X}})$, respectively.
\begin{theorem}\label{cor:KLcontinuity}
	For $p_X=\mathcal{N}(\mu_X,\sigma^2_X)$ and $(R,R_c)\in[0,\infty]^2$, the map  $P\mapsto D(R,R_c,P|\phi_{KL})$ is continuous\footnote{A map $x\mapsto f(x)$ is said to be continuous at $x=\infty$ if $\lim_{x\rightarrow\infty}f(x)=f(\infty)$.} for $P\in[0,\infty]$.
\end{theorem}
\begin{IEEEproof}
	See Appendix \ref{app:KLcontinuity}.	
\end{IEEEproof}




\begin{theorem}\label{cor:W2continuity}
	For $p_X$ with $\mathbb{E}[X^2]<\infty$ and $(R,R_c)\in[0,\infty]^2$, the map $P\mapsto D(R,R_c,P|W^2_2)$ is continuous  for $P\in[0,\infty]$.
\end{theorem}
\begin{IEEEproof}
See Appendix \ref{app:W2continuity}.
\end{IEEEproof}

Now consider the extreme case  $P=\infty$. 
In light of Theorem \ref{thm:outputconstrained}, for $p_{X}$ with $\mathbb{E}[X^2]<\infty$,
\begin{align}
	D(R,R_c,\infty|\phi)&=\inf\limits_{p_{U\hat{X}|X}}\mathbb{E}[(X-\hat{X})^2]\label{eq:DRP}\\
	\mbox{subject to }	&\quad X\leftrightarrow U\leftrightarrow\hat{X}\mbox{ form a Markov chain},\label{eq:constraint1}\\ 
	&\quad I(X;U)\leq R,\label{eq:constraint2}\\
	&\quad I(\hat{X};U)\leq R+R_c,\label{eq:constraint3}
\end{align}
which does not depend on the choice of $\phi$.
Moreover, it can be verified that for $p_X=\mathcal{N}(\mu_X,\sigma^2_X)$,
\begin{align}
		\underline{D}(R,R_c,\infty|\phi_{KL})\nonumber
		&=\underline{D}(R,R_c,\infty|W^2_2)\nonumber\\
		&=\underline{D}'(R,R_c,\infty|W^2_2)\nonumber\\
	&=\sigma^2_Xe^{-2R}\label{eq:underlineD}
\end{align}	
and
\begin{align}	
	\overline{D}(R,R_c,\infty|\phi_{KL})
	&=\overline{D}(R,R_c,\infty|W^2_2)\nonumber\\
	&=\sigma^2_X-\sigma^2_X\xi^2(R,R_c).\label{eq:overlineD}
\end{align} 
Therefore, we shall simply denote $D(R,R_c,\infty|\phi_{KL})$ and $D(R,R_c,\infty|W^2_2)$ by $D(R,R_c,\infty)$, denote $\underline{D}(R,R_c,\infty|\phi_{KL})$, $\underline{D}(R,R_c,\infty|W^2_2)$, and $\underline{D}'(R,R_c,\infty|W^2_2)$ by $\underline{D}(R,R_c,\infty)$, and denote $\overline{D}(R,R_c,\infty|\phi_{KL})$ and
$\overline{D}(R,R_c,\infty|W^2_2)$ by $\overline{D}(R,R_c,\infty)$.

It will be seen that neither $\underline{D}(R,R_c,\infty)$ nor $\overline{D}(R,R_c,\infty)$ is tight in general. This fact can be established by exploiting the connection between rate-distortion-perception coding and entropy-constrained scalar quantization. For $p_X$ with $\mathbb{E}[X^2]<\infty$, let\footnote{The existence of a minimizer for the optimization problem in (\ref{eq:ECSQ}) can be proved via an argument similar to that for Theorem \ref{thm:property}. Here, it suffices to assume $\mathbb{E}[X^2]<\infty$ since the bounded support condition in Theorem \ref{thm:property} is only needed to address the intricacy caused by the Markov chain constraint (\ref{eq:const1}).}
 \begin{align}
	D_e(R,R_c):=\min\limits_{p_{\hat{X}|X}: I(X;\hat{X})\leq R, H(\hat{X})\leq R+R_c}\mathbb{E}[(X-\hat{X})^2],\label{eq:ECSQ}
\end{align}
which is the counterpart of $D(R,R_c,\infty)$ with the decoder restricted to be deterministic \cite[Theorem 5]{SLY15J2}. 
When $R_c=0$, the constraint $I(X;\hat{X})\leq R$ is redundant; as a consequence, 
\begin{align}
	D_e(R,0)=\min\limits_{p_{\hat{X}|X}:  H(\hat{X})\leq R}\mathbb{E}[(X-\hat{X})^2],
\end{align}
which is simply the distortion-rate function for 
entropy-constrained scalar quantization. Note that  
 \begin{align}
 	D_e(R,0)=\min\limits_{p_{\hat{X}}:  H(\hat{X})\leq R}W^2_2(p_X,p_{\hat{X}})
 \end{align}
as every coupling of $p_X$ and $p_{\hat{X}}$ induces a (possibly randomized) scalar quantizer. When $p_{X}$ is absolutely continuous with respect to the Lebesgue measure, $W^2_2(p_X,p_{\hat{X}})$ is attained by a coupling that transforms $p_X$ to $p_{\hat{X}}$ via a determinstic map \cite[Theorem 1.6.2]{PZ20}, 
so there is no loss of optimality in restricting the quantizer  to be deterministic. Moreover, if $p_X$ has a piecewie  monotone and piecewise continuous density, then we can further restrict the deterministic quantizer to be regular \cite[Theorem 5]{GL02}. On the other hand, when $R_c=\infty$, the constraint $H(\hat{X})\leq R+R_c$ is redundant; as a consequence, 
\begin{align}
	D_e(R,\infty)=\min\limits_{p_{\hat{X}|X}:  I(X;\hat{X})\leq R}\mathbb{E}[(X-\hat{X})^2],
\end{align}
which is simply the classical distortion-rate function.
The following result reveals that $D_e(R,R_c)$ is intimately related to $D(R,R_c,\infty)$.








\begin{theorem}\label{thm:connection}
	For $p_X$ with $\mathbb{E}[X^2]<\infty$,
	\begin{align}
		D_e(R,R_c)\geq D(R,R_c,\infty)\geq D_e(R,\infty).\label{eq:order}
	\end{align}
Moreover, if the infimum in (\ref{eq:DRP}) can be attained\footnote{According to Theorem \ref{thm:property}, this assumption holds for $p_X$ with bounded support.}, then
\begin{align}
	D_e(R,R_c)> D_e(R,\infty)\quad\Leftrightarrow\quad D(R,R_c,\infty)>D_e(R,\infty).\label{eq:imply}
\end{align}
\end{theorem}
\begin{IEEEproof}
	See Appendix \ref{app:connection}.
	\end{IEEEproof}

The connection revealed in Theorem \ref{thm:connection} enables us to derive the following result, which indicates that
$\underline{D}(R,R_c,\infty)$ and $\overline{D}(R,R_c,\infty)$ are not tight in general.

\begin{theorem}\label{thm:nontightness}
	For $p_X=\mathcal{N}(\mu_X,\sigma^2_X)$,
	\begin{align}
		D(R,R_c,\infty)>\underline{D}(R,R_c,\infty)\label{eq:Dunderline}
	\end{align}
when $R\in(0,\infty)$ and $R_c\in[0,\infty)$, and 
	\begin{align}
		D(R,R_c,\infty)<\overline{D}(R,R_c,\infty)\label{eq:Doverline}
	\end{align}
when $R_c\in[0,\infty)$ and $R\in(0,\chi(R_c))$, where $\chi(R_c)$ is a positive threshold that depends on $R_c$.
\end{theorem}
\begin{IEEEproof}
	See Appendix \ref{app:nontightness}.	\end{IEEEproof}

\begin{figure}[htbp]
	\centerline{\includegraphics[width=12cm]{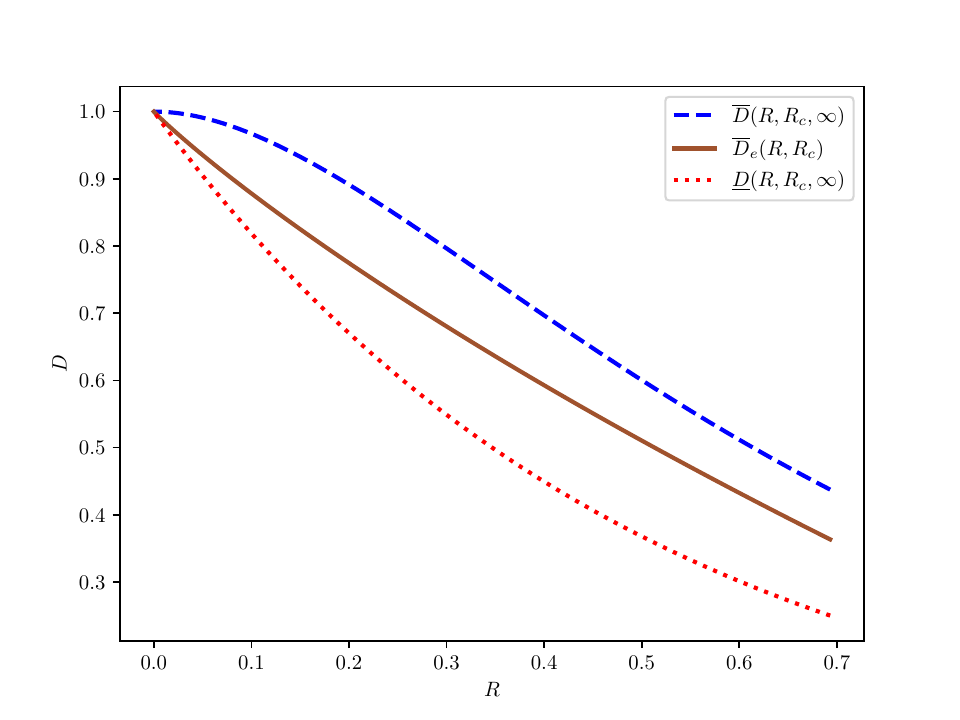}} \caption{Illustrations of $\overline{D}(R,R_c,\infty)$,  $\overline{D}_e(R,R_c)$, and $\underline{D}(R,R_c,\infty)$ for  $p_X=\mathcal{N}(0,1)$ and $R_c=0$.}
	\label{fig:binary_quantization} 
\end{figure}

For $p_{X}=\mathcal{N}(\mu_X,\sigma^2_X)$, we exhibit below an explicit improvement over $\overline{D}(R,0,\infty)$ in the low rate regime. Consider the following binary quantizer:
\begin{align}
	\hat{X}=\begin{cases}
		\mu_X-\frac{\sigma_Xe^{-\frac{\theta^2}{2}}}{\sqrt{2\pi}Q(\theta)}&\mbox{if }\frac{X-\mu_X}{\sigma_X}<\theta,\\
		\mu_X+\frac{\sigma_Xe^{-\frac{\theta^2}{2}}}{\sqrt{2\pi}(1-Q(\theta))}&\mbox{if }\frac{X-\mu_X}{\sigma_X}\geq\theta,
	\end{cases}
\end{align}
where $\theta\geq 0$ and $Q(\theta):=\frac{1}{\sqrt{2\pi}}\int_{-\infty}^{\theta}e^{-\frac{x^2}{2}}\mathrm{d}x$. It can be verified 
that 
\begin{align}
	\mathbb{E}[(X-\hat{X})^2]&=\sigma^2_X-\frac{\sigma^2_Xe^{-\theta^2}}{2\pi Q(\theta)(1-Q(\theta))}\nonumber\\
	&=:D(\theta)
\end{align}	
and
\begin{align}	
	H(\hat{X})&=-Q(\theta)\log Q(\theta)-(1-Q(\theta))\log(1-Q(\theta))\nonumber\\
	&=:R(\theta).
\end{align}
For $R\in(0,\log 2]$, define $\overline{D}_e(R,0)$
via the parametric equations 
$\overline{D}_e(R,0)=D(\theta)$ and $R=R(\theta)$.
Clearly, $\overline{D}_e(R,0)$ is an upper bound on $D_e(R,0)$ and consequently is also an upper bound on $D(R,0,\infty)$ in light of Theorem \ref{thm:connection}.
It can be seen from Fig. \ref{fig:binary_quantization} that $\overline{D}_e(R,0)<\overline{D}(R,0,\infty)$ for $R\in(0,\log 2]$. In particular, we have 
\begin{align}
	\overline{D}_e(\log 2,0)=\frac{\pi-2}{\pi}\sigma^2_X\approx 0.3634\sigma^2_X
\end{align}
while
\begin{align}
	\overline{D}(\log 2, 0,\infty)=\frac{7}{16}\sigma^2_X=0.4375\sigma^2_X.
\end{align}
By contrast, although $\underline{D}(R,R_c,\infty)$ is known to be loose for $R\in(0,\infty)$ and $R_c\in[0,\infty)$, no explicit impprovement has been found (even when $R_c=0$). So $D(R,0,\infty)$ could be situated anywhere between $\overline{D}_e(R,0)$ (inclusive) and $\underline{D}(R,0,\infty)$ (exclusive except at $R=0$) in Fig. \ref{fig:binary_quantization}.

As shown by the following results, Theorem \ref{thm:nontightness} has implications to the weak perception constraint regime in general.

\begin{corollary}\label{cor:weakperceptionKL}
	For $p_X=\mathcal{N}(\mu_X,\sigma^2_X)$, 
	\begin{align}
		D(R,R_c,P|\phi_{KL})>\underline{D}(R,R_c,P|\phi_{KL})\label{eq:KLlower}
	\end{align}
	when $R\in(0,\infty)$, $R_c\in[0,\infty)$, and $P$ is sufficiently large; moreover,
	\begin{align}
		D(R,R_c,P|\phi_{KL})<\overline{D}(R,R_c,P|\phi_{KL}),\label{eq:KLupper}
	\end{align}
	when $R_c\in[0,\infty)$, $R\in(0,\chi(R_c))$, and $P$ is sufficiently large.
\end{corollary}
\begin{IEEEproof}
	See Appendix \ref{app:weakperceptionKL}.
\end{IEEEproof}

\begin{corollary}\label{cor:weakperceptionW2}
	For $p_X=\mathcal{N}(\mu_X,\sigma^2_X)$, 
	\begin{align}
		D(R,R_c,P|W^2_2)>\underline{D}'(R,R_c,P|W^2_2)\label{eq:W2lower}
	\end{align}
when $R\in(0,\infty)$, $R_c\in[0,\infty)$, and $P\in(\gamma'(R,R_c),\infty]$, where $\gamma'(R,R_c)$ is a positive threshold that depends on $(R,R_c)$ and $\gamma'(R,R_c)<P'(R,R_c):=\arg\min\{P\in[0,\infty]:\underline{D}'(R,R_c,P|W^2_2)=\underline{D}'(R,R_c,\infty|W^2_2)\}$\footnote{It can be verified that $P'(R,R_c)=\sigma^2_X(2-e^{-2R}-2\sqrt{(1-e^{-2R})(1-e^{-2(R+R_c)})})$ for $p_X=\mathcal{N}(\mu_X,\sigma^2_X)$.}; moreover,
\begin{align}
	D(R,R_c,P|W^2_2)<\overline{D}(R,R_c,P|W^2_2)\label{eq:W2upper}
\end{align}
when $R_c\in[0,\infty)$, $R\in(0,\chi(R_c))$, and $P\in(\gamma(R,R_c),\infty]$, where $\gamma(R,R_c)$ is a positive threshold that depends on $(R,R_c)$ and $\gamma(R,R_c)< P(R,R_c):=\arg\min\{P\in[0,\infty]:D(R,R_c,P|W^2_2)=D(R,R_c,\infty|W^2_2)\}$\footnote{For $p_X=\mathcal{N}(\mu_X,\sigma^2_X)$, we have $P(R,R_c)>0$ when $R\in[0,\infty)$ and $R_c\in[0,\infty]$
since $D(R,R_c,0|W^2_2)=\overline{D}(R,R_c,0|W^2_2)>\overline{D}(R,R_c,\infty|W^2_2)\geq D(R,R_c,\infty|W^2_2)$.}.
\end{corollary}
\begin{IEEEproof}
	See Appendix \ref{app:weakperceptionW2}.
\end{IEEEproof}

According to \cite[Theorem 2]{XLCZ24}, the upper bounds $\overline{D}(R,R_c,P|\phi_{KL})$ and $\overline{D}(R,R_c,P|W^2_2)$ are tight when the reconstruction distribution $p_{\hat{X}}$ is restricted to be Gaussian. In light of Corollaries \ref{cor:weakperceptionKL} and \ref{cor:weakperceptionW2}, this restriction incurs a penalty in the weak perception constraint regime. In fact, the connection with entropy-constrained scalar quantization suggests that discrete reconstruction distributions might be more preferable in this regime. This is somewhat surprising since $D(R,R_c,P|\phi)$ admits a single-letter characterization, which is typically associated with a Gaussian extremal inequality \cite{GN14}, especially considering the fact that both $\phi_{KL}(p_{\hat{X}}\|p_X)$ and $W^2_2(p_X,p_{\hat{X}})$ favor Gaussian $p_{\hat{X}}$ when $p_{X}$ is a Gaussian distribution (see Lemma \ref{lem:KLW2}).

\section{Conclusion}\label{sec:conclusion}

We have investigated and improved the existing bounds on the quadratic Gaussian distortion-rate-perception function with limited common randomness for the case where the perception measure is given by the Kullback-Leibler divergence or the squared Wasserstein-2 distance. Along the way, a refined version of Talagrand's transportation inequality is established and the connection between rate-distortion-perception coding and entropy-constrained scalar quantization is revealed.

Note that the fundamental rate-distortion-perception tradeoff depends critically on how the perception constraint is formulated. Our work focuses on a particulr formulation where the reconstruction sequence is required to be i.i.d. Therefore, great caution should be executed when utilizing and interpreting the results in the present paper. It is of considerable interest to conduct a comprehensive comparison of different formulations regarding their impacts on the information-theoretic performance limit of rate-distortion-perception coding.

\appendices

\section{Proof of Theorem  \ref{thm:KLW2}}\label{app:KLW2}

We need the following result \cite[Propositions 1 and 2]{XLCZ24} concerning the Gaussian extremal property of the Kullback-Leibler divergence and the squared Wasserstein-2 distance.

\begin{lemma}\label{lem:KLW2}
	For $p_X=\mathcal{N}(\mu_X,\sigma^2_X)$ and $p_{\hat{X}}$ with $\mathbb{E}[\hat{X}^2]<\infty$, 
	\begin{align}
		\phi_{KL}(p_{\hat{X}}\|p_X)
		&\geq\phi_{KL}(p_{\hat{X}^G}\|p_X)\nonumber\\
		&=\log\frac{\sigma_X}{\sigma_{\hat{X}}}+\frac{(\mu_X-\mu_{\hat{X}})^2+\sigma^2_{\hat{X}}-\sigma^2_X}{2\sigma^2_X}		
	\end{align}
	and
	\begin{align}
		W^2_2(p_X,p_{\hat{X}})&\geq W^2_2(p_{X},p_{\hat{X}^G})\nonumber\\
		&=(\mu_X-\mu_{\hat{X}})^2+(\sigma_X-\sigma_{\hat{X}})^2,
	\end{align}
	where $p_{\hat{X}^G}:=\mathcal{N}(\mu_{\hat{X}},\sigma^2_{\hat{X}})$.
\end{lemma}

Lemma \ref{lem:KLW2} indicates that when the reference distribution is Gaussian, replacing the other distribution with its Gaussian counterpart leads to reductions in both the Kullback-Leibler divergence and the squared Wasserstein-2 distance. 
These reductions turn out to be quantitatively related as shown by the next result.

\begin{lemma}\label{lem:general}
	For $p_X=\mathcal{N}(\mu_X,\sigma^2_X)$ and  $p_{\hat{X}}$ with $\mathbb{E}[\hat{X}^2]<\infty$,
	\begin{align}
		&W^2_2(p_X,p_{\hat{X}})-W^2_2(p_X,p_{\hat{X}^G})\leq2\sigma_X\sigma_{\hat{X}}(1-e^{-(\phi_{KL}(p_{\hat{X}}\|p_X)-\phi_{KL}(p_{\hat{X}^G}\|p_X))}).
	\end{align}
\end{lemma}
\begin{IEEEproof}
	Note that
	\begin{align}
		W^2_2(p_X,p_{\hat{X}})	&=(\mu_X-\mu_{\hat{X}})^2+W^2_2(p_{X-\mu_X},p_{\hat{X}-\mu_{\hat{X}}})\nonumber\\
		&=(\mu_X-\mu_{\hat{X}})^2+\sigma^2_XW^2_2(p_{\sigma^{-1}_X(X-\mu_X)}, p_{\sigma^{-1}_X(\hat{X}-\mu_{\hat{X}})})\nonumber\\
		&\stackrel{(a)}{\leq}(\mu_X-\mu_{\hat{X}})^2+\sigma^2_X+\sigma^2_{\hat{X}}-2\sigma^2_X\sqrt{\frac{1}{2\pi e}e^{2h(\sigma^{-1}_X\hat{X})}}\nonumber\\
		&\stackrel{(b)}{=}W^2_2(p_X,p_{\hat{X}^G})+2\sigma_X\sigma_{\hat{X}}-2\sigma^2_X\sqrt{\frac{1}{2\pi e}e^{2h(\sigma^{-1}_X\hat{X})}},\label{eq:rearrange}
	\end{align}
	where ($a$) is due to \cite[Equation (8)]{BWO23} and ($b$) is due to Lemma \ref{lem:KLW2}. Moreover,
	\begin{align}
		h(\sigma^{-1}_X\hat{X})	&=h(\hat{X})-\log\sigma_X\nonumber\\
		&=\frac{1}{2}\log\frac{2\pi e\sigma^2_{\hat{X}}}{\sigma^2_X}-\phi_{KL}(p_{\hat{X}}\|p_X)+\phi_{KL}(p_{\hat{X}^G}\|p_X).\label{eq:deltaKL}
	\end{align}
	Substituting (\ref{eq:deltaKL}) into (\ref{eq:rearrange}) proves Lemma \ref{lem:general}.
\end{IEEEproof}

Now we proceed to prove Theorem \ref{thm:KLW2}. In view of Lemmas \ref{lem:KLW2} and \ref{lem:general}, 
\begin{align}
	W^2_2(p_X,p_{\hat{X}})&\leq\max\limits_{\mu,\sigma}\eta(\mu,\sigma)\label{eq:max1}\\
	\mbox{subject to }&\quad\mu=\mu_X,\label{eq:constr0}\\
	&\quad\sigma\leq\sigma_X,\label{eq:constr1}\\ &\quad\frac{(\mu_X-\mu)^2}{2\sigma^2_X}+\psi(\sigma)\leq\phi_{KL}(p_{\hat{X}}\|p_X),\label{eq:constr2}
\end{align}
where
\begin{align}
	\eta(\mu,\sigma)&:=-2\sigma^2_Xe^{\frac{(\mu_X-\mu)^2+\sigma^2-\sigma^2_X}{2\sigma^2_X}}e^{-\phi_{KL}(p_{\hat{X}}\|p_X)}+(\mu_X-\mu)^2+\sigma^2_X+\sigma^2
\end{align}
and $\psi(\cdot)$ is defined in (\ref{eq:defpsi}).
Since  $\psi(\sigma)$  decreases monotonically from $\infty$ to $0$ as $\sigma$ varies from $0$ to $\sigma_X$ and increases monotonically from $0$ to $\infty$ as $\sigma$ varies from $\sigma_X$ to $\infty$,  there must exist $\underline{\sigma}\leq\sigma_X$ and $\overline{\sigma}\geq\sigma_X$ satisfying
\begin{align}
	\psi(\underline{\sigma})=\psi(\overline{\sigma})=\phi_{KL}(p_{\hat{X}}\|p_X).
\end{align}
Note that (\ref{eq:max1})--(\ref{eq:constr2}) can be written compactly as
\begin{align}
	W^2_2(p_X,p_{\hat{X}})\leq\max\limits_{\sigma\in[\underline{\sigma},\sigma_X]}\eta(\mu_X,\sigma).\label{eq:max1'}
\end{align}
For $\sigma\in[\underline{\sigma},\sigma_X]$,
\begin{align}
	\frac{\partial}{\partial\sigma}\eta(\mu_X,\sigma)&=-2\sigma e^{\frac{\sigma^2-\sigma^2_X}{2\sigma^2_X}}e^{-\phi_{KL}(p_{\hat{X}}\|p_X)}+2\sigma\nonumber\\
	&\geq 0,
\end{align}
which implies the maximum in (\ref{eq:max1'}) is attained at $\sigma=\sigma_X$. So we have
\begin{align}
	W^2_2(p_X,p_{\hat{X}})&\leq\eta(\mu_X,\sigma_X)\nonumber\\
	&=2\sigma^2_X(1-e^{-\phi_{KL}(p_{\hat{X}}\|p_X)}).
\end{align}
This  proves Theorem \ref{thm:KLW2}.


Interestingly,  Talagrand's transportation inequality (\ref{eq:originalTalagrand})
corresponds to the relaxed version without the constraints (\ref{eq:constr0}) and (\ref{eq:constr1}), i.e., 
\begin{align}
	W^2_2(p_X,p_{\hat{X}})&\leq\max\limits_{\mu,\sigma}\eta(\mu,\sigma)\\
	\mbox{subject to }&\quad \frac{(\mu_X-\mu)^2}{2\sigma^2_X}+\psi(\sigma)\leq\phi_{KL}(p_{\hat{X}}\|p_X).\label{eq:constr3}
\end{align}
We now prove this. It can be verified that
\begin{align}
	&\frac{\partial}{\partial(\mu_X-\mu)^2}\eta(\mu,\sigma)=-e^{\frac{(\mu_X-\mu)^2+\sigma^2-\sigma^2_X}{2\sigma^2_X}}e^{-\phi_{KL}(p_{\hat{X}}\|p_X)}+1.
\end{align}
Given $\sigma<\underline{\sigma}$, there is no $\mu$ satisfying (\ref{eq:constr3}).
Given $\sigma\in[\underline{\sigma},\sigma_X]$, for $\mu$ satisfying (\ref{eq:constr3}), we have
\begin{align}
	\frac{\partial}{\partial(\mu_X-\mu)^2}\eta(\mu,\sigma)\geq 0,
\end{align}
which implies that the maximum value of $\eta(\mu,\sigma)$ over $\mu$ satisfying (\ref{eq:constr3}) is attained when
\begin{align}
	\log\frac{\sigma_X}{\sigma}+\frac{(\mu_X-\mu)^2+\sigma^2-\sigma^2_{X}}{2\sigma^2_X}=\phi_{KL}(p_{\hat{X}}\|p_X).
\end{align}
Therefore, for $\sigma\in[\underline{\sigma},\sigma_X]$,
\begin{align}
	\max\limits_{\mu:(\ref{eq:constr3})}\eta(\mu,\sigma)=\kappa(\sigma),
\end{align}
where
\begin{align}
	\kappa(\sigma):=2\sigma^2_X(\phi_{KL}(p_{\hat{X}}\|p_X)-\log\frac{\sigma_X}{\sigma}+1)-2\sigma_X\sigma.
\end{align}
Since the maximum value of $\kappa(\sigma)$ over $\sigma\in[\underline{\sigma},\sigma_X]$ is attained at $\sigma=\sigma_X$, it follows that
\begin{align}
	\max\limits_{\sigma\in[\underline{\sigma},\sigma_X]}\max\limits_{\mu:(\ref{eq:constr3})}\eta(\mu,\sigma)=2\sigma^2_X\phi_{KL}(p_{\hat{X}}\|p_X).\label{eq:comb1}
\end{align}
Given $\sigma\in(\sigma_X,\sqrt{2\sigma^2_X\phi_{KL}(p_{\hat{X}}\|p_X)+\sigma^2_X})$,  for $\mu$ satisfying (\ref{eq:constr3}), we have
\begin{align}
	&\frac{\partial}{\partial(\mu_X-\mu)^2}\eta(\mu,\sigma)\begin{cases}
		\geq 0&\mbox{if }\frac{(\mu_X-\mu)^2+\sigma^2-\sigma^2_{X}}{2\sigma^2_X}\leq\phi_{KL}(p_{\hat{X}}\|p_X),\\
		<0&\mbox{if }\frac{(\mu_X-\mu)^2+\sigma^2-\sigma^2_{X}}{2\sigma^2_X}>\phi_{KL}(p_{\hat{X}}\|p_X),
	\end{cases}	
\end{align}
which implies that the maximum value of $\eta(\mu,\sigma)$ over $\mu$ satisfying (\ref{eq:constr3}) is attained when 
\begin{align}
	\frac{(\mu_X-\mu)^2+\sigma^2-\sigma^2_{X}}{2\sigma^2_X}=\phi_{KL}(p_{\hat{X}}\|p_X).
\end{align}
Therefore, for $\sigma\in(\sigma_X,\sqrt{2\sigma^2_X\phi_{KL}(p_{\hat{X}}\|p_X)+\sigma^2_X})$,
\begin{align}
	\max\limits_{\mu:(\ref{eq:constr3})}\eta(\mu,\sigma)=2\sigma^2_X\phi_{KL}(p_{\hat{X}}\|p_X).
\end{align}
As a consequence,
\begin{align}
	&\max\limits_{\sigma\in(\sigma_X,\sqrt{2\sigma^2_X\phi_{KL}(p_{\hat{X}}\|p_X)+\sigma^2_X})}\max\limits_{\mu:(\ref{eq:constr3})}\eta(\mu,\sigma)=2\sigma^2_X\phi_{KL}(p_{\hat{X}}\|p_X).\label{eq:comb2}
\end{align}
Given $\sigma\in[\sqrt{2\sigma^2_X\phi_{KL}(p_{\hat{X}}\|p_X)+\sigma^2_X},\overline{\sigma}]$,  for $\mu$ satisfying (\ref{eq:constr3}), we have
\begin{align}
	\frac{\partial}{\partial(\mu_X-\mu)^2}\eta(\mu,\sigma)\leq 0,
\end{align}
which implies that the maximum value of $\eta(\mu,\sigma)$ over $\mu$ satisfying (\ref{eq:constr3}) is attained when 
\begin{align}
	(\mu_X-\mu)^2=0,\mbox{ i.e., } \mu=\mu_X.
\end{align}
Therefore, for $\sigma\in[\sqrt{2\sigma^2_X\phi_{KL}(p_{\hat{X}}\|p_X)+\sigma^2_X},\overline{\sigma}]$,
\begin{align}
	\max\limits_{\mu:(\ref{eq:constr3})}\eta(\mu,\sigma)=\kappa'(\sigma),
\end{align}
where
\begin{align}
	\kappa'(\sigma):=-2\sigma^2_Xe^{\frac{\sigma^2-\sigma^2_X}{2\sigma^2_X}}e^{-\phi_{KL}(p_{\hat{X}}\|p_X)}+\sigma^2_X+\sigma^2.
\end{align}
Since the maximum value of $\kappa'(\sigma)$ over $\sigma\in[\sqrt{2\sigma^2_X\phi_{KL}(p_{\hat{X}}\|p_X)+\sigma^2_X},\overline{\sigma}]$ is attained at $\sigma=\sqrt{2\sigma^2_X\phi_{KL}(p_{\hat{X}}\|p_X)+\sigma^2_X}$, it follows that
\begin{align}
	&\max\limits_{\sigma\in[\sqrt{2\sigma^2_X\phi_{KL}(p_{\hat{X}}\|p_X)+\sigma^2_X},\overline{\sigma}]}\max\limits_{\mu:(\ref{eq:constr3})}\eta(\mu,\sigma)=2\sigma^2_X\phi_{KL}(p_{\hat{X}}\|p_X).\label{eq:comb3}
\end{align}
Given $\sigma>\overline{\sigma}$, there is no $\mu$ satisfying (\ref{eq:constr3}).
Combining (\ref{eq:comb1}), (\ref{eq:comb2}), and (\ref{eq:comb3}) proves (\ref{eq:originalTalagrand}).

\section{Proof of Theorem \ref{thm:KLvsW2}}\label{app:nondegenerate}

In view of the definition of $\underline{D}(R,R_c,P|\phi_{KL})$ and $\underline{D}(R,R_c,2\sigma^2_X(1-e^{-P})|W^2_2)$, for the purpose of proving (\ref{eq:W2toKL}), it suffices to show
\begin{align}
	[\sigma(P),\sigma_X]\subseteq[(\sigma_X-\sqrt{2\sigma^2_X(1-e^{-P})})_+,\sigma_X]\label{eq:cond1}
\end{align}
and
\begin{align}
	&\sigma^2_{\hat{X}}-(\sigma_Xe^{-(R+R_c)}-\sqrt{2\sigma^2_X(1-e^{-P})})_+^2\geq\sigma^2_{\hat{X}}-\sigma^2_{\hat{X}}e^{-2(R+R_c+P-\psi(\sigma_{\hat{X}}))}\label{eq:cond2}
\end{align}
for $\sigma_{\hat{X}}\in[\sigma(P),\sigma_X]$.
Invoking (\ref{eq:refinedTalagrand}) with $p_{\hat{X}}=\mathcal{N}(\mu_X,\sigma(P))$ (see also Lemma \ref{lem:KLW2} for the expressions of the Kullback-Leibler divergence and the squared Wasserstein-2 distance between two Gaussian distributions)
\begin{align}
	(\sigma_X-\sigma(P))^2\leq 2\sigma^2_X(1-e^{-P}),
\end{align}
from which (\ref{eq:cond1}) follows immediately.
Note that (\ref{eq:cond2}) is trivially true when 	
$e^{-(R+R_c)}\leq\sqrt{2(1-e^{-P})}$. When $e^{-(R+R_c)}>\sqrt{2(1-e^{-P})}$, it can be written equivalently as
\begin{align}
	\sqrt{2(1-e^{-P})}\geq e^{-(R+R_c)}(1-e^{-(P+\frac{\sigma^2_X-\sigma^2_{\hat{X}}}{2\sigma^2_X})}).
\end{align}
Since $e^{-(R+R_c)}\leq 1$ and
\begin{align}
	1-e^{-(P+\frac{\sigma^2_X-\sigma^2_{\hat{X}}}{2\sigma^2_X})}\leq1-e^{-(P+\frac{\sigma^2_X-\sigma^2(P)}{2\sigma^2_X})}
\end{align}
for $\sigma_{\hat{X}}\in[\sigma(P),\sigma_X]$, it suffices to show
\begin{align}
	\sqrt{2(1-e^{-P})}\geq1-e^{-(P+\frac{\sigma^2_X-\sigma^2(P)}{2\sigma^2_X})}.\label{eq:suffshow}
\end{align}
According to the definition of $\sigma(P)$,
\begin{align}
	P=\log\frac{\sigma_X}{\sigma(P)}+\frac{\sigma^2(P)-\sigma^2_X}{2\sigma^2_X}.\label{eq:sigmaP}
\end{align}
Substituting (\ref{eq:sigmaP}) into (\ref{eq:suffshow}) gives
\begin{align}
	\sqrt{2(1-e^{\log \frac{\sigma(P)}{\sigma_X}-\frac{\sigma^2(P)}{2\sigma^2_X}+\frac{1}{2}})}\geq 1-\frac{\sigma(P)}{\sigma_X}. \label{eq:rewrite}
\end{align}
We can rewrite (\ref{eq:rewrite}) as
\begin{align}
	\tau(\beta)\geq 0,\label{eq:tau}
\end{align}
where 
\begin{align}
	\tau(\beta):=1-2\beta e^{-\frac{\beta^2}{2}+\frac{1}{2}}+2\beta-\beta^2
\end{align}
with $\beta:=\frac{\sigma(P)}{\sigma_X}$.
Note that	$\beta\in[0,1]$. We have
\begin{align}
	\frac{\mathrm{d}\tau(\beta)}{\mathrm{d}\beta}&=-2e^{-\frac{\beta^2}{2}+\frac{1}{2}}+2\beta^2e^{-\frac{\beta^2}{2}+\frac{1}{2}}+2-2\beta\nonumber\\
	&\leq -2(1-\beta^2)+2-2\beta\nonumber\\
	&=-2(1-\beta)\beta\nonumber\\
	&\leq 0.
\end{align}
Since $\tau(1)=0$, it follows that $\tau(\beta)\geq 0$ for $\beta\in[0,1]$, which verifies (\ref{eq:tau}) and consequently proves (\ref{eq:cond2}).

Now we proceed to prove (\ref{KLtoW2}), which is equivalent to 
\begin{align}
	\overline{D}(R,R_c,2\sigma^2_X(1-e^{-P})|W^2_2)\leq\overline{D}(R,R_c,P|\phi_{KL}).
\end{align}
Since $\overline{D}(R,R_c,P|\phi_{KL})=\overline{D}(R,R_c,(\sigma_X-\sigma(P))^2|W^2_2)$, it suffices to show
\begin{align}
	(\sigma_X-\sigma(P))^2\leq 2\sigma^2_X(1-e^{-P}),
\end{align}
i.e.,
\begin{align}
	P\geq\log\frac{2\sigma^2_X}{\sigma^2_X-\sigma^2(P)+2\sigma_X\sigma(P)}.\label{eq:tosub}
\end{align}
Substituting (\ref{eq:sigmaP}) into (\ref{eq:tosub}) and rearranging the inequality yields
\begin{align}
	\log\frac{\sigma^2_X-\sigma^2(P)+2\sigma_X\sigma(P)}{2\sigma_X\sigma(P)}\geq\frac{\sigma^2_X-\sigma^2(P)}{2\sigma^2_X},
\end{align}
which is indeed true since 
\begin{align}
\log\frac{\sigma^2_X-\sigma^2(P)+2\sigma_X\sigma(P)}{2\sigma_X\sigma(P)}	&\stackrel{(a)}{\geq} 1-\frac{2\sigma_X\sigma(P)}{\sigma^2_X-\sigma^2(P)+2\sigma_X\sigma(P)}\nonumber\\
	&=\frac{\sigma^2_X-\sigma^2(P)}{\sigma^2_X-\sigma^2(P)+2\sigma_X\sigma(P)}\nonumber\\
	&\geq\frac{\sigma^2_X-\sigma^2(P)}{2\sigma^2_X},
\end{align}
where ($a$) is due to 
$\log z\geq 1-\frac{1}{z}$ for $z>0$. 
This completes the proof of  (\ref{KLtoW2}).

\section{Proof of Theorem \ref{thm:improvedlowerbound}}\label{app:improvedlowerbound}

It is known \cite[Remark 2 and Lemma 3]{XLCZ24} that
\begin{align}
	D(R,R_c,P|W^2_2)\geq&\inf\limits_{p_{\hat{X}}}\sigma^2_X+\sigma^2_{\hat{X}}-2\sigma_X\sqrt{(1-e^{-2R})(\sigma^2_{\hat{X}}-D(R+R_c|p_{\hat{X}}))}\\
	\mbox{subject to}&\quad \mu_{\hat{X}}=\mu_X,\label{eq:mean}\\
	&\quad\sigma_{\hat{X}}\leq\sigma_X,\label{eq:variance}\\
	&\quad W^2_2(p_X,p_{\hat{X}})\leq P,\label{eq:W2}
\end{align}
where
\begin{align}
	D(R+R_c|p_{\hat{X}}):=\inf\limits_{p_{\hat{Y}|\hat{X}}:I(\hat{X};\hat{Y})\leq R+R_c}\mathbb{E}[(\hat{X}-\hat{Y})^2].
\end{align}
In light of Lemma \ref{lem:KLW2}, the constraints (\ref{eq:mean})--(\ref{eq:W2}) imply $\sigma_{\hat{X}}\in[(\sigma_X-\sqrt{P})_+,\sigma_X]$. The following result provides a lower bound on $D(R+R_c|p_{\hat{X}})$ and proves
\begin{align}
	&D(R,R_c,P|W^2_2)\geq\inf\limits_{\sigma_{\hat{X}}\in[(\sigma_X-\sqrt{P})_+,\sigma_X]}\sup\limits_{\alpha>0} \sigma^2_X+\sigma^2_{\hat{X}}-2\sigma_X\sqrt{(1-e^{-2R})(\sigma^2_{\hat{X}}-\delta^2_+(\sigma_{\hat{X}},\alpha))}.\label{eq:infsup}
\end{align}

\begin{lemma}\label{lem:GaussianW2}
	For $p_X=\mathcal{N}(\mu_X,\sigma^2_X)$ and  $p_{\hat{X}}$ with $W^2_2(p_X,p_{\hat{X}})\leq P$, 
	\begin{align}
		D(R+R_c|p_{\hat{X}})\geq\sup\limits_{\alpha>0}	\frac{(\sigma_Xe^{-(R+R_c)}-G(\alpha))^2_+}{\alpha^2},
	\end{align}	
	where
	\begin{align}
		G(\alpha):=\sqrt{\sigma^2_X-\alpha((\mu_X-\mu_{\hat{X}})^2+\sigma^2_X+\sigma^2_{\hat{X}}-P)+\alpha^2\sigma^2_{\hat{X}}}.
	\end{align}
\end{lemma}
\begin{IEEEproof}
	First let $p_X$ and $p_{\hat{X}}$ be coupled according to the joint distribution attaining $W^2_2(p_X,p_{\hat{X}})$.
	Then add $\hat{Y}$ into the probability space  such that $X\leftrightarrow\hat{X}\leftrightarrow\hat{Y}$ form a Markov chain and $I(\hat{X};\hat{Y})\leq R+R_c$.  For any $\alpha>0$,
	\begin{align}
		&\mathbb{E}[((X-\mu_X)-\alpha(\hat{Y}-\mu_{\hat{Y}}))^2]\nonumber\\
		&=\mathbb{E}[((X-\mu_X)-\alpha(\hat{X}-\mu_{\hat{X}}))^2]+\alpha^2\mathbb{E}[((\hat{X}-\mu_{\hat{X}})-(\hat{Y}-\mu_{\hat{Y}}))^2]+2\alpha\mathbb{E}[((X-\mu_X)-(\hat{X}-\mu_{\hat{X}}))((\hat{X}-\mu_{\hat{X}})-(\hat{Y}-\mu_{\hat{Y}}))]\nonumber\\
		&\leq(\sqrt{\mathbb{E}[((X-\mu_X)-\alpha(\hat{X}-\mu_{\hat{X}}))^2]}+\alpha\sqrt{\mathbb{E}[((\hat{X}-\mu_{\hat{X}})-(\hat{Y}-\mu_{\hat{Y}}))^2]})^2\nonumber\\
		&\leq(\sqrt{\mathbb{E}[((X-\mu_X)-\alpha(\hat{X}-\mu_{\hat{X}}))^2]}+\alpha\sqrt{\mathbb{E}[(\hat{X}-\hat{Y})^2]})^2\nonumber\\
		&=(\sqrt{\sigma^2_X-2\alpha\rho\sigma_X\sigma_{\hat{X}}+\alpha^2\sigma^2_{\hat{X}}}+\alpha\sqrt{\mathbb{E}[(\hat{X}-\hat{Y})^2]})^2,\label{eq:hand1}
	\end{align}
where  $\rho$ denotes the correlation coefficient of $X$ and $\hat{X}$.	On the other hand,
	\begin{align}
		\mathbb{E}[((X-\mu_X)-\alpha(\hat{Y}-\mu_{\hat{Y}}))^2]&\stackrel{(a)}{\geq}\sigma^2_Xe^{-2I(X-\mu_X;\alpha(\hat{Y}-\mu_{\hat{Y}}))}\nonumber\\
		&=\sigma^2_Xe^{-2I(X;\hat{Y})}\nonumber\\
		&\stackrel{(b)}{\geq}\sigma^2_Xe^{-2I(\hat{X};\hat{Y})}\nonumber\\
		&\geq\sigma^2_Xe^{-2(R+R_c)},\label{eq:hand2}
	\end{align}
	where ($a$) and ($b$) are due to the Shannon lower bound \cite[Equation (13.159)]{CT91} and the data processing inequality \cite[Theorem 2.8.1]{CT91}, respectively. 
	Combining (\ref{eq:hand1}) and (\ref{eq:hand2}) yields
	\begin{align}
		&\mathbb{E}[(\hat{X}-\hat{Y})^2]\geq\frac{(\sigma_Xe^{-(R+R_c)}-\sqrt{\sigma^2_X-2\alpha\rho\sigma_X\sigma_{\hat{X}}+\alpha^2\sigma^2_{\hat{X}}})^2_+}{\alpha^2}.\label{eq:rho}
	\end{align}
	It can be verified that 
	\begin{align}
		P&\geq W^2_2(p_X,p_{\hat{X}})\nonumber\\
		&=\mathbb{E}[(X-\hat{X})^2]\nonumber\\
		&=(\mu_X-\mu_{\hat{X}})^2+\mathbb{E}[((X-\mu_X)-(\hat{X}-\mu_{\hat{X}}))^2]\nonumber\\
		&=(\mu_X-\mu_{\hat{X}})^2+\sigma^2_X-2\rho\sigma_X\sigma_{\hat{X}}+\sigma^2_{\hat{X}},
	\end{align}
	which implies
	\begin{align}
		2\rho\sigma_X\sigma_{\hat{X}}\geq(\mu_X-\mu_{\hat{X}})^2+\sigma^2_X+\sigma^2_{\hat{X}}-P.\label{eq:subrho}
	\end{align}
	Substituting (\ref{eq:subrho}) into (\ref{eq:rho}) proves Lemma \ref{lem:GaussianW2}.
\end{IEEEproof}

To establish the first inequality in (\ref{eq:twoinequalities}), we shall demonstrate that ``$\inf$" in (\ref{eq:infsup}) can be replaced by ``$\min$". It suffices to consider the case $R\in(0,\infty)$ and $R_c\in[0,\infty)$ since otherwise the infimum is clearly attainable. 
The problem boils down to showing that the map $\sigma_{\hat{X}}\mapsto\sup_{\alpha>0}\delta_+(\sigma_{\hat{X}},\alpha)$ is continuous for $\sigma_{\hat{X}}\in[(\sigma_X-\sqrt{P})_+,P]$.

Obviously, $\sup_{\alpha>0}\delta_+(\sigma_{\hat{X}},\alpha)=0$ if and only if $\sup_{\alpha>0}\delta(\sigma_{\hat{X}},\alpha)\leq 0$, where
\begin{align}
	&\delta(\sigma_{\hat{X}},\alpha):=\frac{\sigma_Xe^{-(R+R_c)}-\sqrt{\sigma^2_X-\alpha(\sigma^2_X+\sigma^2_{\hat{X}}-P)+\alpha^2\sigma^2_{\hat{X}}}}{\alpha}.
\end{align}
Note that $\sup_{\alpha>0}\delta(\sigma_{\hat{X}},\alpha)\leq 0$ is equivalent to
\begin{align}
	P\geq\sup\limits_{\alpha>0}\sigma^2_X+\sigma^2_{\hat{X}}-\frac{\sigma^2_X(1-e^{-2(R+R_c)})}{\alpha}-\alpha\sigma^2_{\hat{X}}.\label{eq:rewrittennew}
\end{align}
Since
\begin{align}
	&\sup\limits_{\alpha>0}\sigma^2_X+\sigma^2_{\hat{X}}-\frac{\sigma^2_X(1-e^{-2(R+R_c)})}{\alpha}-\alpha\sigma^2_{\hat{X}}=\sigma^2_X+\sigma^2_{\hat{X}}-2\sigma_X\sigma_{\hat{X}}\sqrt{1-e^{-2(R+R_c)}},
\end{align}
one can rewrite (\ref{eq:rewrittennew}) as
\begin{align}
	P\geq\sigma^2_X+\sigma^2_{\hat{X}}-2\sigma_X\sigma_{\hat{X}}\sqrt{1-e^{-2(R+R_c)}}.\label{eq:becomes1new}
\end{align}

On the other hand, we have
$\sup_{\alpha>0}\delta_+(\sigma_{\hat{X}},\alpha)=\sup_{\alpha>0}\delta(\sigma_{\hat{X}},\alpha)>0$ when 
\begin{align}
	P<\sigma^2_X+\sigma^2_{\hat{X}}-2\sigma_X\sigma_{\hat{X}}\sqrt{1-e^{-2(R+R_c)}}.\label{eq:Plessthan}
\end{align}
If $\sigma_{\hat{X}}=\sigma_X-\sqrt{P}$, then 
\begin{align}
	\delta(\sigma_{\hat{X}},\alpha)=\frac{\sigma_Xe^{-(R+R_c)}-|\sigma_X-\alpha\sigma_{\hat{X}}|}{\alpha}
\end{align}
and $\sup_{\alpha>0}\delta(\sigma_{\hat{X}},\alpha)$ is attained at\footnote{It follows by $\sigma_{\hat{X}}=\sigma_X-\sqrt{P}$ and (\ref{eq:Plessthan}) that $\sigma_{\hat{X}}>0$.} 
\begin{align}
	\alpha=\frac{\sigma_X}{\sigma_{\hat{X}}}.\label{eq:degenerate}
\end{align} 
If $\sigma_{\hat{X}}>\sigma_X-\sqrt{P}$, then
\begin{align}
	\sigma^2_X-\alpha(\sigma^2_X+\sigma^2_{\hat{X}}-P)+\alpha^2\sigma^2_{\hat{X}}>0
\end{align}
for $\alpha>0$.
As shown below,  $\frac{\partial}{\partial\alpha}\delta(\sigma_{\hat{X}},\alpha)=0$ has a unique solution, denoted as $\hat{\alpha}$, for $\alpha>0$. 
It can be verified that
\begin{align}
	&\frac{\partial}{\partial\alpha}\delta(\sigma_{\hat{X}},\alpha)=\frac{\alpha(\sigma^2_X+\sigma^2_{\hat{X}}-P-2\alpha\sigma^2_{\hat{X}})}{2\alpha^2\sqrt{\sigma^2_{X}-\alpha(\sigma^2_{X}+\sigma^2_{\hat{X}}-P)+\alpha^2\sigma^2_{\hat{X}}}}-\frac{\sigma_Xe^{-(R+R_c)}-\sqrt{\sigma^2_X-\alpha(\sigma^2_X+\sigma^2_{\hat{X}}-P)+\alpha^2\sigma^2_{\hat{X}}}}{\alpha^2}.
\end{align}
Setting $\frac{\partial}{\partial\alpha}\delta(\sigma_{\hat{X}},\alpha)=0$ gives
\begin{align}
	&2\sigma^2_X-\alpha(\sigma^2_X+\sigma^2_{\hat{X}}-P)=2\sigma_Xe^{-(R+R_c)}\sqrt{\sigma^2_X-\alpha(\sigma^2_X+\sigma^2_{\hat{X}}-P)+\alpha^2\sigma^2_{\hat{X}}}.\label{eq:equ}
\end{align}
Note that (\ref{eq:equ}) has a solution  in $(0,\frac{2\sigma^2_X}{\sigma^2_X+\sigma^2_{\hat{X}}-P})$ since its left-hand side is greater than its right-hand side   when $\alpha=0$ and is less than its right-hand side  when $\alpha=\frac{2\sigma^2_X}{\sigma^2_X+\sigma^2_{\hat{X}}-P}$.  By taking the square of both sides of  (\ref{eq:equ}) and simplifying the expression, we get
\begin{align}
	&\alpha^2((\sigma^2_X+\sigma^2_{\hat{X}}-P)^2-4\sigma^2_X\sigma^2_{\hat{X}}e^{-2(R+R_c)})-4\alpha\sigma^2_X(\sigma^2_X+\sigma^2_{\hat{X}}-P)(1-e^{-2(R+R_c)})+4\sigma^4_X(1-e^{-2(R+R_c)})=0.\label{eq:equ2}
\end{align} 
If $\sigma^2_X+\sigma^2_{\hat{X}}-P=2\sigma_X\sigma_{\hat{X}}e^{-(R+R_c)}$, then (\ref{eq:equ2}) has only one solution, given by 
\begin{align}
	\hat{\alpha}:=\frac{\sigma^2_X}{\sigma^2_X+\sigma^2_{\hat{X}}-P},\label{eq:linear}
\end{align}
which is also the unique solution to $\frac{\partial}{\partial\alpha}\delta(\sigma_{\hat{X}},\alpha)=0$.
If $\sigma^2_X+\sigma^2_{\hat{X}}-P<2\sigma_X\sigma_{\hat{X}}e^{-(R+R_c)}$, then (\ref{eq:equ2}) has two solutions with different signs and only the positive one, given by
\begin{align}
	\hat{\alpha}&:=\frac{2\sigma^2_X(\sigma^2_X+\sigma^2_{\hat{X}}-P)(1-e^{-2(R+R_c)})-2\sigma^2_Xe^{-(R+R_c)}\sqrt{(4\sigma^2_X\sigma^2_{\hat{X}}-(\sigma^2_X+\sigma^2_{\hat{X}}-P)^2)(1-e^{-2(R+R_c)})}}{(\sigma^2_X+\sigma^2_{\hat{X}}-P)^2-4\sigma^2_X\sigma^2_{\hat{X}}e^{-2(R+R_c)}},\label{eq:also}
\end{align}
is the solution to $\frac{\partial}{\partial\alpha}\delta(\sigma_{\hat{X}},\alpha)=0$ for $\alpha>0$. If $\sigma^2_X+\sigma^2_{\hat{X}}-P>2\sigma_X\sigma_{\hat{X}}e^{-(R+R_c)}$, then (\ref{eq:equ2}) has two positive solutions and only the small one, also given by (\ref{eq:also}), is the solution to $\frac{\partial}{\partial\alpha}\delta(\sigma_{\hat{X}},\alpha)=0$.
Indeed, the large one is a solution to the following equation obtained by negating the left-hand side of (\ref{eq:equ}):
\begin{align}
	&\alpha(\sigma^2_X+\sigma^2_{\hat{X}}-P)-2\sigma^2_X=2\sigma_Xe^{-(R+R_c)}\sqrt{\sigma^2_X-\alpha(\sigma^2_X+\sigma^2_{\hat{X}}-P)+\alpha^2\sigma^2_{\hat{X}}}.\label{eq:equopp}
\end{align}
This can be verified by noticing that (\ref{eq:equopp}) has a solution in $(\frac{2\sigma^2_X}{\sigma^2_X+\sigma^2_{\hat{X}}-P},\infty)$ since its left-hand side is less than its right-hand side when $\alpha=\frac{2\sigma^2_X}{\sigma^2_X+\sigma^2_{\hat{X}}-P}$ and is greater than its right-hand side when $\alpha$ is sufficiently large. For $\sigma_{\hat{X}}$ satisfying $\sigma^2_X+\sigma^2_{\hat{X}}-P=2\sigma_X\sigma_{\hat{X}}e^{-(R+R_c)}$, we have
\begin{align}
	&\lim\limits_{\epsilon\rightarrow0}\frac{2\sigma^2_X(\sigma^2_X+\sigma^2_{\hat{X}}(\epsilon)-P)(1-e^{-2(R+R_c)})-2\sigma^2_Xe^{-(R+R_c)}\sqrt{(4\sigma^2_X\sigma^2_{\hat{X}}(\epsilon)-(\sigma^2_X+\sigma^2_{\hat{X}}(\epsilon)-P)^2)(1-e^{-2(R+R_c)})}}{(\sigma^2_X+\sigma^2_{\hat{X}}(\epsilon)-P)^2-4\sigma^2_X\sigma^2_{\hat{X}}(\epsilon)e^{-2(R+R_c)}}\nonumber\\
	&=\frac{\sigma^2_X}{\sigma^2_X+\sigma^2_{\hat{X}}-P},
\end{align}
where $\sigma_{\hat{X}}(\epsilon)=\sigma_{\hat{X}}+\epsilon$.
Moreover, setting $\sigma_{\hat{X}}=\sigma_X-\sqrt{P}$ in (\ref{eq:also}) gives $\hat{\alpha}=\frac{\sigma_X}{\sigma_{\hat{X}}}$.
Therefore,  (\ref{eq:degenerate}) and (\ref{eq:linear}) can be viewed as the degenerate versions of (\ref{eq:also}). 
Since $\delta(\sigma_{\hat{X}},\alpha)<0$ when $\alpha$ is either close to zero from the positive side or sufficiently large,  $\hat{\alpha}$ must be the unique maximizer of both $\delta(\sigma_{\hat{X}},\alpha)$ and $\delta_+(\sigma_{\hat{X}},\alpha)$
for $\alpha>0$.
This implies the continuity of $\sigma_{\hat{X}}\mapsto\sup_{\alpha>0}\delta_+(\sigma_{\hat{X}},\alpha)$ for $\sigma_{\hat{X}}$ over the region defined  by (\ref{eq:Plessthan}).

It remains to show that $\delta_+(\sigma_{\hat{X}},\hat{\alpha})\rightarrow 0$ as $\sigma_{\hat{X}}$, confined to the region defined by (\ref{eq:Plessthan}), converges to some $\sigma$ satisfying $P=\sigma^2_X+\sigma^2-2\sigma_X\sigma\sqrt{1-e^{-2(R+R_c)}}$. First consider the scenario where $\sigma=0$, which implies $P=\sigma^2_X$. We have $\hat{\alpha}\rightarrow\infty$ as $\sigma_{\hat{X}}\rightarrow 0$, and consequently
\begin{align}
	\lim\limits_{\sigma_{\hat{X}}\rightarrow0}\delta_+(\sigma_{\hat{X}},\hat{\alpha})
	&=\lim\limits_{\sigma_{\hat{X}}\rightarrow0}\left(\frac{\sigma_Xe^{-(R+R_c)}}{\alpha}-\sqrt{\frac{\sigma^2_X}{\alpha^2}-\frac{\sigma^2_X+\sigma^2_{\hat{X}}-P}{\alpha}+\sigma^2_{\hat{X}}}\right)_+\nonumber\\
	&=0.
\end{align}
Next consider the scenario where $\sigma>0$. We have $\hat{\alpha}\rightarrow\frac{\sigma^2_X+\sigma^2-P}{2\sigma^2}$ as $\sigma_{\hat{X}}\rightarrow\sigma$, and consequently
\begin{align}
	\lim\limits_{\sigma_{\hat{X}}\rightarrow0}\delta_+(\sigma_{\hat{X}},\hat{\alpha})&=\delta_+\left(\sigma,\frac{\sigma^2_X+\sigma^2-P}{2\sigma^2}\right)\nonumber\\
	&=0.
\end{align}
This completes the proof of the first inequality in (\ref{eq:twoinequalities}).





The second inequality in (\ref{eq:twoinequalities}) follows from the fact that 
\begin{align}
	&\underline{D}(R,R_c,P|W^2_2)=\min\limits_{\sigma_{\hat{X}}\in[(\sigma_X-\sqrt{P})_+,\sigma_X]} \sigma^2_X+\sigma^2_{\hat{X}}-2\sigma_X\sqrt{(1-e^{-2R})(\sigma^2_{\hat{X}}-\delta^2_+(\sigma_{\hat{X}},1))}.\label{eq:infimum}
\end{align}
Now we proceed to identify the sufficient and necessary condition under which this inequality is strict. It suffices to consider the case $R\in(0,\infty)$ and $R_c\in[0,\infty)$ since otherwise $\underline{D}'(R,R_c,P|W^2_2)$ clearly coincides with $\underline{D}(R,R_c,P|W^2_2)$. 
Note  that the minimum in (\ref{eq:infimum}) is attained at and only at \cite[Appendix F]{XLCZ24}
\begin{align}
	\sigma_{\hat{X}}&=\hat{\sigma}:=\begin{cases}
		\sigma_X\sqrt{1-e^{-2R}}&\mbox{if }\frac{\sqrt{P}}{\sigma_X}\geq(1-\sqrt{1-e^{-2R}})\vee e^{-(R+R_c)},\\
		\sigma_X-\sqrt{P}
		&\mbox{if }\frac{\sqrt{P}}{\sigma_X}\in[e^{-(R+R_c)},1-\sqrt{1-e^{-2R}}),\\
		\sqrt{\sigma^2_X(1-e^{-2R})+(\sigma_Xe^{-(R+R_c)}-\sqrt{P})^2}&\mbox{if }\frac{\sqrt{P}}{\sigma_X}\in[\nu(R,R_c),e^{-(R+R_c)}),\\
		\sigma_X-\sqrt{P}&\mbox{if }\frac{\sqrt{P}}{\sigma_X}<\nu(R,R_c)\wedge e^{-(R+R_c)},
	\end{cases}
\end{align}
where
\begin{align}
	\nu(R,R_c):=\frac{e^{-2R}-e^{-2(R+R_c)}}{2-2e^{-(R+R_c)}}.\label{eq:nu}
\end{align}
We have the following observation:
$\underline{D}'(R,R_c,P|W^2_2)>\underline{D}(R,R_c,P|W^2_2)$ if and only if
\begin{align}
	&\sup\limits_{\alpha>0}\sigma^2_X+\hat{\sigma}^2-2\sigma_X\sqrt{(1-e^{-2R})(\hat{\sigma}^2-\delta^2_+(\hat{\sigma},\alpha))}>\sigma^2_X+\hat{\sigma}^2-2\sigma_X\sqrt{(1-e^{-2R})(\hat{\sigma}^2-\delta^2_+(\hat{\sigma},1))}.\label{eq:iff}
\end{align}

\underline{``If" part}: Assume the minimum in (\ref{eq:minsup}) is attained at $\sigma_{\hat{X}}=\tilde{\sigma}$. If $\tilde{\sigma}=\hat{\sigma}$, we have
\begin{align}
	\underline{D}'(R,R_c,P|W^2_2)
	&=\sup\limits_{\alpha>0}\sigma^2_X+\hat{\sigma}^2-2\sigma_X\sqrt{(1-e^{-2R})(\hat{\sigma}^2-\delta^2_+(\hat{\sigma},\alpha))}\nonumber\\
	&>\sigma^2_X+\hat{\sigma}^2-2\sigma_X\sqrt{(1-e^{-2R})(\hat{\sigma}^2-\delta^2_+(\hat{\sigma},1))}\nonumber\\
	&=\underline{D}(R,R_c,P|W^2_2).
\end{align}
If $\tilde{\sigma}\neq\hat{\sigma}$, we have
\begin{align}
	\underline{D}'(R,R_c,P|W^2_2)
	&=\sup\limits_{\alpha>0}\sigma^2_X+\tilde{\sigma}^2-2\sigma_X\sqrt{(1-e^{-2R})(\tilde{\sigma}^2-\delta^2_+(\tilde{\sigma},\alpha))}\nonumber\\
	&\geq\sigma^2_X+\tilde{\sigma}^2-2\sigma_X\sqrt{(1-e^{-2R})(\tilde{\sigma}^2-\delta^2_+(\tilde{\sigma},1))}\nonumber\\
	&\stackrel{(a)}{>}\sigma^2_X+\hat{\sigma}^2-2\sigma_X\sqrt{(1-e^{-2R})(\hat{\sigma}^2-\delta^2_+(\hat{\sigma},1))}\nonumber\\
	&=\underline{D}(R,R_c,P|W^2_2),
\end{align}
where ($a$) is due to the fact that $\hat{\sigma}$ is the unique minimizer of (\ref{eq:infimum}). Thus, $\underline{D}'(R,R_c,P|W^2_2)>\underline{D}(R,R_c,P|W^2_2)$ holds either way.

\underline{``Only if" part}: This is because
\begin{align}
	\sup\limits_{\alpha>0}\sigma^2_X+\hat{\sigma}^2-2\sigma_X\sqrt{(1-e^{-2R})(\hat{\sigma}^2-\delta^2_+(\hat{\sigma},\alpha))}
	&\geq\underline{D}'(R,R_c,P|W^2_2)\nonumber\\
	&>\underline{D}(R,R_c,P|W^2_2)\nonumber\\
	&=\sigma^2_X+\hat{\sigma}^2-2\sigma_X\sqrt{(1-e^{-2R})(\hat{\sigma}^2-\delta^2_+(\hat{\sigma},1))}.
\end{align}


Equipped with the above observation, we shall treat the following two cases separately.

1) $P\geq\sigma^2_Xe^{-2(R+R_c)}$: In this case, $\delta_+(\sigma_{\hat{X}},1)=0$. Therefore, (\ref{eq:iff}) holds if and only  if $\sup_{\alpha>0}\delta_+(\hat{\sigma},\alpha)>0$, which, in light of (\ref{eq:Plessthan}), is equivalent to
\begin{align}
	P<\sigma^2_X+\hat{\sigma}^2-2\sigma_X\hat{\sigma}\sqrt{1-e^{-2(R+R_c)}}.\label{eq:becomes1}
\end{align}

For the subcase $\frac{\sqrt{P}}{\sigma_X}\geq(1-\sqrt{1-e^{-2R}})\vee e^{-(R+R_c)}$, we have $\hat{\sigma}=\sigma_X\sqrt{1-e^{-2R}}$, and consequently (\ref{eq:becomes1}) becomes
\begin{align*}
	P<\sigma^2_X(2-e^{-2R}-2\sqrt{(1-e^{-2R})(1-e^{-2(R+R_c)})}).
\end{align*}
For the subcase $\frac{\sqrt{P}}{\sigma_X}\in[e^{-(R+R_c)},1-\sqrt{1-e^{-2R}})$, we have $\hat{\sigma}=\sigma_X-\sqrt{P}$,  and consequently (\ref{eq:becomes1}) becomes 
\begin{align}
	0<(2\sigma^2_X-2\sigma_X\sqrt{P})(1-\sqrt{1-e^{-2(R+R_c)}}),
\end{align}
which holds trivially. Combining the analyses for these two subcases shows that $\frac{P}{\sigma^2_X}$ must fall into the following interval:
\begin{align}
	[e^{-2(R+R_c)},2-e^{-2R}-2\sqrt{(1-e^{-2R})(1-e^{-2(R+R_c)})}).\label{eq:interval}
\end{align}
Note that
\begin{align}
	e^{-2(R+R_c)}
	&=\left.2-e^{-2R}-\frac{1-e^{-2(R+R_c)}}{\alpha}-\alpha(1-e^{-2R})\right|_{\alpha=1}\nonumber\\
	&\leq\sup\limits_{\alpha>0}2-e^{-2R}-\frac{1-e^{-2(R+R_c)}}{\alpha}-\alpha(1-e^{-2R})\nonumber\\
	&=2-e^{-2R}-2\sqrt{(1-e^{-2R})(1-e^{-2(R+R_c)})},
\end{align}
where the supremum is attained at and only at $\alpha=\sqrt{\frac{1-e^{-2(R+R_c)}}{1-e^{-2R}}}$. 
Thus, the interval in (\ref{eq:interval}) is nonempty unless $R_c=0$.
	
2) $P<\sigma^2_Xe^{-2(R+R_c)}$: In this case, $\delta_+(\sigma_{\hat{X}},1)>0$. 
Clearly, $\delta_+(\sigma_{\hat{X}},\alpha)=\delta(\sigma_{\hat{X}},\alpha)$ whenever $\delta_+(\sigma_{\hat{X}},\alpha)>0$. Since $\delta_+(\sigma_{\hat{X}},1)>0$, we must have $\delta_+(\sigma_{\hat{X}},\alpha)=\delta(\sigma_{\hat{X}},\alpha)$ in a neighbourhood of $\alpha=1$.
Setting $\left.\frac{\partial}{\partial\alpha}\delta(\sigma_{\hat{X}},\alpha)\right|_{\alpha=1}=0$ gives
\begin{align}
	\sigma_{\hat{X}}=\sigma^*_{\hat{X}}:=\sqrt{\sigma^2_X-2\sigma_Xe^{-(R+R_c)}\sqrt{P}+P}.\label{eq:critical}
\end{align}

For the subcase $\frac{\sqrt{P}}{\sigma_X}\in[\nu(R,R_c),e^{-(R+R_c)})$, we have 
$\hat{\sigma}=\sqrt{\sigma^2_X(1-e^{-2R}+e^{-2(R+R_c)})-2\sigma_Xe^{-(R+R_c)}\sqrt{P}+P}$, which, in view of (\ref{eq:critical}), implies $\left.\frac{\partial}{\partial\alpha}\delta(\hat{\sigma},\alpha)\right|_{\alpha=1}\neq 0$ unless $R_c=0$. Note that in this subcase, $P=0\Rightarrow\nu(R,R_c)=0\Rightarrow R_c=0$. 
For the subcase $\frac{\sqrt{P}}{\sigma_X}<\nu(R,R_c)\wedge e^{-(R+R_c)}$, we have
$\hat{\sigma}=\sqrt{\sigma^2_X-2\sigma_X\sqrt{P}+P}$, which, in view of (\ref{eq:critical}), implies $\left.\frac{\partial}{\partial\alpha}\delta(\hat{\sigma},\alpha)\right|_{\alpha=1}\neq 0$ unless $P=0$. Note that this subcase is void when $R_c=0$ since $\nu(R,0)=0$. Combining the analyses for these two subcases shows that if $R_c>0$ and $P>0$, then $\left.\frac{\partial}{\partial\alpha}\delta(\hat{\sigma},\alpha)\right|_{\alpha=1}\neq 0$, which further implies (\ref{eq:iff}).

It remains to show $\left.\hat{\alpha}\right|_{\sigma_{\hat{X}}=\hat{\sigma}}=1$ when $R_c=0$ or $P=0$. This can be accomplished via a direct verification. The proof of Theorem \ref{thm:improvedlowerbound} is thus complete.

\section{Proof of (\ref{eq:Rthreshold})}\label{app:Rthreshold}

Clearly, (\ref{eq:Pthreshold}) holds for all $R\geq 0$ when $P\geq\sigma^2_X$. It remains to consider the case $P\in(0,\sigma^2_X)$. We can write  (\ref{eq:Pthreshold}) equivalently as
\begin{align}
	2-z-\frac{P}{\sigma^2_X}\leq 2\sqrt{(1-z)(1-ze^{-2R_c})},\label{eq:rew}
\end{align}
where $z:=e^{-2R}$.
Note that 
\begin{align}
	2-z-\frac{P}{\sigma^2_X}= 2\sqrt{(1-z)(1-ze^{-2R_c})}\label{eq:squarez}
\end{align}
has a solution in $(0,1)$ since its left-hand side is less than its right-hand side when $z=0$ and is greater than its right-hand side when $z=1$. By taking the square of both sides of (\ref{eq:squarez}) and simplifying the expression, we get
\begin{align}
	\zeta_1z^2-\zeta_2z+\zeta_3=0.\label{eq:zeta}
\end{align}
It is easy to see that $\zeta_2>0$ and $\zeta_3>0$.
If $\zeta_1=0$ (i.e., $R_c=\log 2$), then  (\ref{eq:zeta}) has only one solution, given by
\begin{align}
	\hat{z}:=\frac{\zeta_3}{\zeta_2},
\end{align}
which is also the unique solution to (\ref{eq:squarez}). If $\zeta_1<0$ (i.e., $R_c>\log 2$), then  (\ref{eq:zeta}) has two solutions with different signs and only the positive one, given by
\begin{align}
	\hat{z}:=\frac{\zeta_2-\sqrt{\zeta^2_2-4\zeta_1\zeta_3}}{2\zeta_1},\label{eq:alsozeta}
\end{align}
is the solution to (\ref{eq:squarez}) for $z\in(0,1)$. If $\zeta_1>0$ (i.e, $R_c<\log 2$), then (\ref{eq:zeta}) has two positive solutions and only the small one, also given by (\ref{eq:alsozeta}), is the solution to  (\ref{eq:squarez}). Indeed, the large one is a solution to the following equation obtained  by negating the left-hand side of  (\ref{eq:squarez}):
\begin{align}
	-2+z+\frac{P}{\sigma^2_X}= 2\sqrt{(1-z)(1-ze^{-2R_c})}.\label{eq:negeq}
\end{align}
This can be verified by noticing that (\ref{eq:negeq}) has a solution in $(1,\infty)$ since its left-hand side is less than its right-hand side when $z=1$ and is greater than its right-hand side when $z$ is sufficiently large. Therefore, $\hat{z}$ is the unique solution to (\ref{eq:rew}) for $z\in(0,1)$, and consequently 
(\ref{eq:rew}) holds if and only if $z\in[0,\hat{z}]$, i.e., $R\geq-\frac{1}{2}\log\hat{z}$.

\section{Proof of Theorem \ref{thm:property}}\label{app:property}

\begin{lemma}\label{lem:tightness}
	For the optimization problem in  (\ref{eq:inf}), there is no loss of generality in assuming $U=\mathbb{E}[X|U]$ almost surely and $\mathbb{E}[\hat{X}^2]\leq(1+\sqrt{2})^2\mathbb{E}[X^2]$.
\end{lemma}	
\begin{IEEEproof}
	Note that
	\begin{align}
		D(R,R_c,P|\phi)\leq 2\mathbb{E}[X^2]
	\end{align} 
	since we can trivially let $X$, $U$, and $\hat{X}$ be mutually independent and $p_{\hat{X}}=p_X$. Therefore, it suffices to consider $p_{U\hat{X}|X}$ with $\mathbb{E}[\hat{X}^2]\leq (1+\sqrt{2})^2\mathbb{E}[X^2]$ because otherwise
	\begin{align}
		\mathbb{E}[(X-\hat{X})^2]&=\mathbb{E}[X^2]+\mathbb{E}[\hat{X}^2]-2\mathbb{E}[X\hat{X}]\nonumber\\
		&\geq\mathbb{E}[X^2]+\mathbb{E}[\hat{X}^2]-2\sqrt{\mathbb{E}[X^2]\mathbb{E}[\hat{X}^2]}\nonumber\\
		&>\mathbb{E}[X^2]+(1+\sqrt{2})^2\mathbb{E}[X^2]-2(1+\sqrt{2})\mathbb{E}[X^2]\nonumber\\
		&=2\mathbb{E}[X^2].
	\end{align}

	For any $p_{U\hat{X}|X}$ satisfying (\ref{eq:const1})--(\ref{eq:const4}), let $\hat{U}:=\mathbb{E}[X|U]$. Construct $p_{U'\hat{X}'|X}$ such that 	
	$X\leftrightarrow U'\leftrightarrow \hat{X}'$ form a Markov chain, $p_{U'|X}=p_{\hat{U}|X}$, and $p_{\hat{X}'|U'}=p_{\hat{X}|\hat{U}}$. Clearly, 
	\begin{align}
		&I(X;U')=I(X;\hat{U})\stackrel{(a)}{\leq} I(X;U)\leq R,\\
		&I(\hat{X}';U')=I(\hat{X};\hat{U})\stackrel{(b)}{\leq} I(\hat{X};U)\leq R+R_c,		
	\end{align}
	where ($a$) and ($b$) are due to the data processing inequality \cite[Theorem 2.8.1]{CT91}. Moreover, we have $p_{\hat{X}'}=p_{\hat{X}}$ and consequently
	\begin{align}
		\phi(p_{X},p_{\hat{X}'})=\phi(p_{X},p_{\hat{X}})\leq P.
	\end{align}
	It can also be verified that
	\begin{align}
		\mathbb{E}[(X-\hat{X}')]&\stackrel{(c)}{=}\mathbb{E}[(X-U')^2]+\mathbb{E}[(\hat{X}'-U')^2]\nonumber\\
		&=\mathbb{E}[(X-\hat{U})^2]+\mathbb{E}[(\hat{X}-\hat{U})^2]\nonumber\\
		&\stackrel{(d)}{=}\mathbb{E}[(X-\hat{X})],\label{eq:similar}
	\end{align}
where ($c$) and ($d$) follow  respectively from the facts that  $U'=\mathbb{E}[X|U',\hat{X}']$ and $\hat{U}=\mathbb{E}[X|\hat{U},\hat{X}]$ almost surely.
Therefore, there is no loss of optimality in replacing $p_{U\hat{X}|X}$ with $p_{U'\hat{X}'|X}$.
\end{IEEEproof}

Now we proceed to prove Theorem \ref{thm:property}. For any positive integer $k$, in light of Lemma \ref{lem:tightness}, there exists $p_{U^{(k)}\hat{X}^{(k)}|X}$ satisfying
\begin{align}
	&I(X;U^{(k)})\leq R,\label{eq:constra1}\\
	&I(\hat{X}^{(k)};U^{(k)})\leq R+R_c,\label{eq:constra2}\\
	&\phi(p_X,p_{\hat{X}^{(k)}})\leq P,\label{eq:constra3}\\
	&U^{(k)}=\mathbb{E}[X|U^{(k)}]\mbox{ almost surely},\label{eq:as_k}\\
	&\mathbb{E}[(\hat{X}^{(k)})^2]\leq (1+\sqrt{2})^2\mathbb{E}[X^2]
\end{align}
as well as the Markov chain constraint $X\leftrightarrow U^{(k)}\leftrightarrow\hat{X}^{(k)}$
such that
\begin{align}
	\mathbb{E}[(X-\hat{X}^{(k)})^2]\leq D(R,R_c,P|\phi)+\frac{1}{k}.\label{eq:combin1}
\end{align}
The sequence $\{p_{XU^{(k)}\hat{X}^{(k)}}\}_{k=1}^{\infty}$ is tight \cite[Definition in Appendix II]{GN14} since given any $\epsilon>0$,
\begin{align}
	&\mathbb{P}\Big\{X^2\leq \frac{3}{\epsilon}\mathbb{E}[X^2], (U^{(k)})^2\leq \frac{3}{\epsilon}\mathbb{E}[X^2], (\hat{X}^{(k)})^2\leq \frac{3(1+\sqrt{2})^2}{\epsilon}\mathbb{E}[X^2]\Big\}\nonumber\\
	&\geq 1-\mathbb{P}\Big\{X^2>\frac{3}{\epsilon}\mathbb{E}[X^2]\}-\mathbb{P}\{(U^{(k)})^2>\frac{3}{\epsilon}\mathbb{E}[X^2]\Big\}-\mathbb{P}\Big\{(\hat{X}^{(k)})^2>\frac{3(1+\sqrt{2})^2}{\epsilon}\mathbb{E}[X^2]\Big\}\nonumber\\
	&\geq 1-\frac{\epsilon}{3}-\frac{\mathbb{E}[(U^{(k)})^2]\epsilon}{3\mathbb{E}[X^2]}-\frac{\mathbb{E}[(\hat{X}^{(k)})^2]\epsilon}{3(1+\sqrt{2})^2\mathbb{E}[X^2]}\nonumber\\
	&\geq1-\epsilon
\end{align}
for all $k$. By Prokhorov's theorem \cite[Theorem 4]{GN14}, there exists a subsequence $\{p_{XU^{(k_m)}\hat{X}^{(k_m)}}\}_{m=1}^{\infty}$ converging weakly to some distriution $p_{XU^*\hat{X}^*}$. 
Since $p_X$ has bounded support, it follows by \cite[Theorem 3]{WV12} that
\begin{align}
	\mathbb{E}[(X-\mathbb{E}[X|U^*,\hat{X}^*])^2]\geq\limsup\limits_{m\rightarrow\infty}\mathbb{E}[(X-\mathbb{E}[X|U^{(k_m)},\hat{X}^{k_m}])^2]=\limsup\limits_{m\rightarrow\infty}\mathbb{E}[(X-U^{(k_m)})^2].\label{eq:mmse1}
\end{align}
On the other hand,  as the map $(x,u)\mapsto(x-u)^2$ is continuous and bounded from below, we have
\begin{align}
	\mathbb{E}[(X-U^*)^2]\leq\liminf\limits_{m\rightarrow\infty}\mathbb{E}[(X-U^{(k_m)})^2],\label{eq:mmse2}
\end{align}
which, together with (\ref{eq:mmse1}), implies
\begin{align}
U^*=\mathbb{E}[X|U^*,\hat{X}^*]\mbox{ almost surely}.\label{eq:as}
\end{align}  Moreover, by the lower semicontinuity of mutual information and $p_{\hat{X}}\mapsto\phi(p_X, p_{\hat{X}})$ in the topology of weak convergence,
\begin{align}
	&I(X;U^*)\leq\liminf\limits_{m\rightarrow\infty}I(X;U^{(k_m)}),\label{eq:wc1}\\
	&I(\hat{X}^*;U^*)\leq\liminf\limits_{m\rightarrow\infty}I(\hat{X}^{(k_m)};U^{(k_m)}),\label{eq:wc2}\\
	&\phi(p_X,p_{\hat{X}^*})\leq\liminf\limits_{m\rightarrow\infty}\phi(p_X,p_{\hat{X}^{(k_m)}}).\label{eq:wc3}
\end{align}

Construct $p_{U'\hat{X}'|X}$ such that 	
$X\leftrightarrow U'\leftrightarrow \hat{X}'$ form a Markov chain, $p_{U'|X}=p_{U^*|X}$, and $p_{\hat{X}'|U'}=p_{\hat{X}^*|U^*}$. In view of (\ref{eq:constra1})--(\ref{eq:constra3}) and (\ref{eq:wc1})--(\ref{eq:wc3}),
\begin{align}
	&I(X;U')=I(X;U^*)\leq R,\\
	&I(\hat{X}';U')=I(\hat{X}^*;U^*)\leq R+R_c,\\
	&\phi(p_X,p_{\hat{X}'})=\phi(p_X,p_{\hat{X}^*})\leq P.
\end{align}
Similarly to (\ref{eq:similar}), we have
\begin{align}
	\mathbb{E}[(X-\hat{X}')^2]&=\mathbb{E}[(X-U')^2]+\mathbb{E}[(\hat{X}'-U')]\nonumber\\
	&=\mathbb{E}[(X-U^*)^2]+\mathbb{E}[(\hat{X}^*-U^*)]\nonumber\\
	&=\mathbb{E}[(X-\hat{X}^*)^2].\label{eq:combin2}
\end{align}
Since the map $(x,\hat{x})\mapsto(x-\hat{x})^2$ is continuous and bounded from below, it follows that
\begin{align}	
	&\mathbb{E}[(X-\hat{X}^*)^2]\leq\liminf\limits_{m\rightarrow\infty}\mathbb{E}[(X-\hat{X}^{(k_m)})^2.\label{eq:combin3}
\end{align} 
Combining (\ref{eq:combin1}), (\ref{eq:combin2}), and (\ref{eq:combin3}) shows
\begin{align}	
	&\mathbb{E}[(X-\hat{X}')^2]\leq D(R,R_c,P|\phi).
\end{align} 
Therefore, the infimum in (\ref{eq:inf}) is attained
at $p_{U'\hat{X}'|X}$.




The above argument can be easily leveraged to prove  the lower semicontinuity of $(R,R_c,P)\mapsto D(R,R_c,P|\phi)$, which implies the desired right-continuity property since the map $(R,R_c,P)\mapsto D(R,R_c,P|\phi)$ is monotonically decreasing in each of its variables.

The following subtlety in this proof is noteworthy. It is tempting to claim that the weak convergence limit  $p_{XU^*\hat{X}^*}$ automatically satisfies the Markov chain constraint $X\leftrightarrow U^*\leftrightarrow \hat{X}^*$. We are unable to confirm this claim. In fact, this claim is false if (\ref{eq:as_k}) does not hold. For example, let $U^{(k)}:=\frac{1}{k}U$ and
\begin{align}
	X=\hat{X}^{(k)}:=\begin{cases}
		1&\mbox{if } U\geq 0,\\
		-1&\mbox{if } U<0,
		\end{cases}
\end{align}
where $U$ is a standard Gaussian random variable. It is clear that $X\leftrightarrow U^{(k)}\leftrightarrow \hat{X}^{(k)}$ form a Markov chain for any positive integer $k$. However, the Markov chain constraint  is violated by the weak convergence limit $p_{XU^*\hat{X}^*}$ since $X$ and $\hat{X}^*$ are two identical symmetric Bernoulli random variables whereas $U^*$ is a constant zero. Our key observation is that it suffices to have (\ref{eq:as}), with which the Markov chain structure can be  restored without affecting the end-to-end distortion (see the construction of $p_{U'\hat{X}'|X}$). Nevertheless, we only manage to establish (\ref{eq:as}) when $p_X$ has bounded support. Note that, according to the exampel above, the minimum mean square error is not necessarily preserved under weak convergence if  (\ref{eq:as_k}) does not hold. Indeed, while $\mathbb{E}[(X-\mathbb{E}[X|U^{(k)}])^2]=0$ for any positive integer $k$, we have $\mathbb{E}[(X-\mathbb{E}[X|U^*])^2]=1$.


\section{Proof of Theorem  \ref{cor:KLcontinuity}}\label{app:KLcontinuity}

We need the following well-known result regarding the Ornstein-Uhlenbeck flow (see, e.g., \cite[Lemma 1]{WJ18}).

\begin{lemma}\label{lem:KLcontinuity}
	For $p_X=\mathcal{N}(\mu_X,\sigma^2_X)$ and $p_{\hat{X}}$ with $\mu_{\hat{X}}=\mu_X$ and $\mathbb{E}[\hat{X}^2]<\infty$, 
	let $\hat{X}(\lambda):=\mu_X+\sqrt{1-\lambda}(\hat{X}-\mu_X)+\sqrt{\lambda}(\bar{X}-\mu_X)$, where $\bar{X}$ is independent of $\hat{X}$ and has the same distribution as $X$.		
	The map $\lambda\mapsto\phi_{KL}(p_{\hat{X}(\lambda)}\|p_X)$ is continuous\footnote{$\phi_{KL}(p_{\hat{X}(\lambda)}\|p_X)$ varies continuously from $\phi_{KL}(p_{\hat{X}}\|p_X)$ to $0$ as $\lambda$ increases from $0$ to $1$. Note that $\phi_{KL}(p_{\hat{X}(\lambda)}\|p_X)<\infty$ for $\lambda\in(0,1]$ and
		 $\lim_{\lambda\rightarrow 0}\phi_{KL}(p_{\hat{X}}(\lambda)\|p_X)=\phi_{KL}(p_{\hat{X}}\|p_X)$ even if $\phi_{KL}(p_{\hat{X}}\|p_X)=\infty$ (in this sense, the map $\lambda\mapsto\phi_{KL}(p_{\hat{X}(\lambda)}\|p_X)$ is continuous at $\lambda=0$).}, decreasing, and convex for $\lambda\in[0,1]$.
\end{lemma}

Now we proceed to prove Theorem \ref{cor:KLcontinuity}.  Given $\epsilon>0$, there exists $p_{U\hat{X}|X}$ satisfying (\ref{eq:const1})--(\ref{eq:const4}) with $\phi=\phi_{KL}$ and $\mathbb{E}[(X-\hat{X})^2]\leq D(R,R_c,P|\phi_{KL})+\epsilon$. Without loss of generality, we assume $\mu_{\hat{X}}=\mu_X$ and $\sigma_{\hat{X}}\leq\sigma_X$ \cite[Lemma 1]{XLCZ24}. For $\lambda\in[0,1]$, let $\hat{X}(\lambda):=\mu_X+\sqrt{1-\lambda}(\hat{X}-\mu_X)+\sqrt{\lambda}(\bar{X}-\mu_X)$, where $\bar{X}$ is assumed to be independent of $(X,U,\hat{X})$ and have the same distribution as $X$. Note that $X\leftrightarrow U\leftrightarrow\hat{X}\leftrightarrow \hat{X}(\lambda)$ form a Markov chain.
By the data processing inequality \cite[Theorem 2.8.1]{CT91} and  (\ref{eq:const3}), 
\begin{align*}
	I(\hat{X}(\lambda);U)\leq I(\hat{X};U)\leq R+R_c.
\end{align*}

First consider the case $P\in(0,\infty)$. In light of Lemma \ref{lem:KLcontinuity}, given $\tilde{P}\in(0,P]$,  there exists $\tilde{\lambda}\in[0,1]$ such that $\phi_{KL}(p_{\hat{X}(\tilde{\lambda})}\|p_X)=\phi_{KL}(p_{\hat{X}}\|p_X)\wedge\tilde{P}$; moreover, we have\footnote{When $\phi_{KL}(p_{\hat{X}}\|p_X)=0$, we can set $\tilde{\lambda}=0$ and consequently $\tilde{\lambda}\leq\frac{P-\tilde{P}}{\tilde{P}}$ still holds.}
\begin{align}
	\tilde{\lambda}\leq 1-\frac{\phi_{KL}(p_{\hat{X}}\|p_X)\wedge \tilde{P}}{\phi_{KL}(p_{\hat{X}}\|p_X)}\leq\frac{P-\tilde{P}}{\tilde{P}}.\label{eq:tildelambda}
\end{align}
It can be verified that 
\begin{align}
	&D(R,R_c,\tilde{P}|\phi_{KL})-D(R,R_c,P|\phi_{KL})\nonumber\\
	&\leq D(R,R_c,\tilde{P}|\phi_{KL})-\mathbb{E}[(X-\hat{X})^2]+\epsilon\nonumber\\
	&\leq\mathbb{E}[(X-\hat{X}(\tilde{\lambda}))^2]-\mathbb{E}[(X-\hat{X})^2]+\epsilon\nonumber\\
	&=2\mathbb{E}[(X-\hat{X})(\hat{X}-\hat{X}(\tilde{\lambda}))]+\mathbb{E}[(\hat{X}-\hat{X}(\tilde{\lambda}))^2]+\epsilon\nonumber\\
	&\leq2\sqrt{\mathbb{E}[(X-\hat{X})^2]\mathbb{E}[(\hat{X}-\hat{X}(\tilde{\lambda}))^2]}+\mathbb{E}[(\hat{X}-\hat{X}(\tilde{\lambda}))^2]+\epsilon\nonumber\\
	&=2\sqrt{\mathbb{E}[(X-\hat{X})^2]((1-\sqrt{1-\tilde{\lambda}})^2\sigma^2_{\hat{X}}+\tilde{\lambda}\sigma^2_X)}+(1-\sqrt{1-\tilde{\lambda}})^2\sigma^2_{\hat{X}}+\tilde{\lambda}\sigma^2_X+\epsilon\nonumber\\
	&\leq(4\sqrt{2-2\sqrt{1-\tilde{\lambda}}}+2-2\sqrt{1-\tilde{\lambda}})\sigma^2_X+\epsilon\nonumber\\
	&\leq(4\sqrt{2-2\sqrt{\frac{(2\tilde{P}-P)_+}{\tilde{P}}}}+2-2\sqrt{\frac{(2\tilde{P}-P)_+}{\tilde{P}}})\sigma^2_X+\epsilon.\label{eq:hold}
\end{align}
This proves
\begin{align}
	D(R,R_c,\tilde{P}|\phi_{KL})-D(R,R_c,P|\phi_{KL}) \leq(4\sqrt{2-2\sqrt{\frac{(2\tilde{P}-P)_+}{\tilde{P}}}}+2-2\sqrt{\frac{(2\tilde{P}-P)_+}{\tilde{P}}})\sigma^2_X.\label{eq:imply1}
\end{align}

Next consider the case $P=\infty$.  In light of Lemma \ref{lem:KLcontinuity}, given $\tilde{P}\in[0,\infty)$, there exists $\tilde{\lambda}\in(0,1]$ such that $\phi_{KL}(p_{\hat{X}(\tilde{\lambda})}\|p_X)\leq \tilde{P}$; moreover, we can require $\tilde{\lambda}\rightarrow 0$ as $\tilde{P}\rightarrow\infty$. It can be verified that  
\begin{align}
	\lim\limits_{\tilde{P}\rightarrow\infty}D(R,R_c,\tilde{P}|\phi_{KL})&\leq\lim\limits_{\tilde{\lambda}\rightarrow 0}\mathbb{E}[(X-\hat{X}(\lambda))^2]\nonumber\\
	&=\mathbb{E}[(X-\hat{X})^2]\nonumber\\
	&\leq D(R,R_c,\infty|\phi_{KL})+\epsilon.
\end{align}
This proves
\begin{align}
		\lim\limits_{\tilde{P}\rightarrow\infty}D(R,R_c,\tilde{P}|\phi_{KL})\leq D(R,R_c,\infty|\phi_{KL}).\label{eq:imply2}
\end{align}

Finally, consider the case $P=0$. Given $\tilde{P}\in[0,\infty]$ and $\epsilon>0$, there exists $p_{U\hat{X}|X}$ 
satisfying (\ref{eq:const1})--(\ref{eq:const3}), $\phi_{KL}(p_{\hat{X}}\|p_X)\leq\tilde{P}$, and $\mathbb{E}[(X-\hat{X})^2]\leq D(R,R_c,\tilde{P}|\phi_{KL})+\epsilon$. Without loss of generality, we assume $\mu_{\hat{X}}=\mu_X$ and $\sigma_{\hat{X}}\leq\sigma_X$ \cite[Lemma 1]{XLCZ24}. Let $\bar{X}'$ be jointly distributed with $(X,U,\hat{X})$
such that $X\leftrightarrow U\leftrightarrow\hat{X}\leftrightarrow\bar{X}'$ form a Markov chain, $\bar{X}'\sim p_X$, and $\mathbb{E}[(\bar{X}'-\hat{X})^2]=W^2_2(p_X,p_{\hat{X}})$. By the data processing inequality \cite[Theorem 2.8.1]{CT91} and  (\ref{eq:const3}), 
\begin{align*}
	I(\bar{X}';U)\leq I(\hat{X};U)\leq R+R_c.
\end{align*}
It can be verified that
\begin{align}
	&D(R,R_c,0|\phi_{KL})-D(R,R_c,\tilde{P}|\phi_{KL})\nonumber\\
	&\leq D(R,R_c,0|\phi_{KL})-\mathbb{E}[(X-\hat{X})^2]+\epsilon\nonumber\\
	&\leq\mathbb{E}[(X-\bar{X}')^2]-\mathbb{E}[(X-\hat{X})^2]+\epsilon\nonumber\\
	&=2\mathbb{E}[(X-\hat{X})(\hat{X}-\bar{X}')]+\mathbb{E}[(\hat{X}-\bar{X}')^2]+\epsilon\nonumber\\
	&\leq 2\sqrt{\mathbb{E}[(X-\hat{X})^2]\mathbb{E}[(\hat{X}-\bar{X}')^2]}+\mathbb{E}[(\hat{X}-\bar{X}')^2]+\epsilon\nonumber\\
	&=2\sqrt{\mathbb{E}[(X-\hat{X})^2]W^2_2(p_X,p_{\hat{X}})}+W^2_2(p_X,p_{\hat{X}})+\epsilon\nonumber\\
	&\stackrel{(a)}{\leq}2\sqrt{2\sigma^2_X\mathbb{E}[(X-\hat{X})^2]\phi_{KL}(p_{\hat{X}}\|p_X)}+2\sigma^2_X\phi_{KL}(p_{\hat{X}}\|p_X)+\epsilon\nonumber\\
	&\leq2\sqrt{2\sigma^2_X\mathbb{E}[(X-\hat{X})^2]\tilde{P}}+2\sigma^2_X\tilde{P}+\epsilon\nonumber\\
&\leq(4{\sqrt{2\tilde{P}}}+2\tilde{P})\sigma^2_X+\epsilon,
\end{align}
where ($a$) is due to Talagrand's transportation  inequality \cite{Talagrand96}. 
This proves
\begin{align}
	D(R,R_c,0|\phi_{KL})-D(R,R_c,\tilde{P}|\phi_{KL})\leq (4{\sqrt{2\tilde{P}}}+2\tilde{P})\sigma^2_X.\label{eq:imply3}
\end{align}

In view of (\ref{eq:imply1}), (\ref{eq:imply2}), and (\ref{eq:imply3}),  the desired continuity property follows by the fact that the map $P\mapsto D(R,R_c,P|W^2_2)$ is monotonically decreasing.

\section{Proof of Theorem  \ref{cor:W2continuity}}\label{app:W2continuity}

We need the following result \cite[Theorem 3]{FMM21} regarding the distortion-perception tradeoff in the quadratic 
Wasserstein space.

\begin{lemma}\label{lem:DPtradeoff}
	For $\lambda\in[0,1]$ and $p_{X\hat{X}}$ with $\mathbb{E}[X^2]<\infty$ and $\mathbb{E}[\hat{X}^2]<\infty$, let
$\hat{X}(\lambda):=(1-\lambda)\tilde{X}+\lambda\bar{X}$, where $\tilde{X}:=\mathbb{E}[X|\hat{X}]$, and $\bar{X}$ is
jointly distributed with $(X,\hat{X})$ such that $X\leftrightarrow\hat{X}\leftrightarrow\bar{X}$ form a Markov chain, $\bar{X}\sim p_{X}$, and $\mathbb{E}[(\bar{X}-\tilde{X})^2]=W^2_2(p_{X},p_{\tilde{X}})$. We have
\begin{align}
	&\mathbb{E}[(X-\hat{X}(\lambda))^2]=\mathbb{E}[(X-\tilde{X})^2]+(W_2(p_{X},p_{\tilde{X}})-W_2(p_X,p_{\hat{X}(\lambda)}))^2\label{eq:tight}
\end{align}
and
\begin{align}
	W^2_2(p_X,p_{\hat{X}(\lambda)})=(1-\lambda)^2W^2_2(p_X,p_{\tilde{X}}).\label{eq:W2expression}
\end{align}
Moreover, for any $\hat{X}'$ jointly distributed with $(X,\hat{X})$ such that $X\leftrightarrow\hat{X}\leftrightarrow\hat{X}'$ form a Markov chain and $\mathbb{E}[(\hat{X}')^2]<\infty$,
\begin{align}
	&\mathbb{E}[(X-\hat{X}')^2]\geq\mathbb{E}[(X-\tilde{X})^2]+(W_2(p_{X},p_{\tilde{X}})-W_2(p_X,p_{\hat{X}'}))^2_+.\label{eq:bound}
\end{align}
\end{lemma}


Now we proceed to prove Theorem \ref{cor:W2continuity}. Given $\epsilon>0$, there exists  $p_{U\hat{X}|X}$ satisfying (\ref{eq:const1})--(\ref{eq:const4}) with $\phi=W^2_2$ and $\mathbb{E}[(X-\hat{X})^2]\leq D(R,R_c,P|W^2_2)+\epsilon$. Let $\bar{X}$ be jointly distributed with $(X,U,\hat{X})$ such that $(X,U)\leftrightarrow\hat{X}\leftrightarrow\bar{X}$ form a Markov chain, $\bar{X}\sim p_{X}$, and $\mathbb{E}[(\bar{X}-\tilde{X})^2]=W^2_2(p_{X},p_{\tilde{X}})$, where
$\tilde{X}:=\mathbb{E}[X|\hat{X}]$. Moreover, let 
$\hat{X}(\lambda):=(1-\lambda)\tilde{X}+\lambda\bar{X}$ for $\lambda\in[0,1]$. Note that $X\leftrightarrow U\leftrightarrow\hat{X}\leftrightarrow \hat{X}(\lambda)$ form a Markov chain. By the data processing inequality \cite[Theorem 2.8.1]{CT91} and (\ref{eq:const3}), 
\begin{align}
	I(\hat{X}(\lambda);U)\leq I(\hat{X};U)\leq R+R_c.
\end{align}

Given $\tilde{P}\in[0,P]$, in light of (\ref{eq:W2expression}) in Lemma \ref{lem:DPtradeoff}, there exists $\tilde{\lambda}\in[0,1]$ such that $W^2_2(p_X,p_{\hat{X}(\tilde{\lambda})})=W^2_2(p_X,p_{\tilde{X}})\wedge \tilde{P}$. We have
\begin{align}
&D(R,R_c,\tilde{P}|W^2_2)-D(R,R_c,P|W^2_2)\nonumber\\
&\leq D(R,R_c,\tilde{P}|W^2_2)-\mathbb{E}[(X-\hat{X})^2]+\epsilon\nonumber\\	&\leq\mathbb{E}[(X-\hat{X}(\tilde{\lambda}))^2]-\mathbb{E}[(X-\hat{X})^2]+\epsilon\nonumber\\
	&\stackrel{(a)}{\leq}(W_2(p_{X},p_{\tilde{X}})-W_2(p_X,p_{\hat{X}(\tilde{\lambda})}))^2-(W_2(p_{X},p_{\tilde{X}})-W_2(p_X,p_{\hat{X}}))^2_++\epsilon\nonumber\\
	&\leq(W_2(p_{X},p_{\tilde{X}})-W_2(p_X,p_{\hat{X}(\tilde{\lambda})}))^2-(W_2(p_{X},p_{\tilde{X}})-(W_2(p_X,p_{\tilde{X}})\wedge P))^2+\epsilon\nonumber\\
	&=(2W_2(p_{X},p_{\tilde{X}})-(W_2(p_X,p_{\tilde{X}})\wedge P)-W_2(p_X,p_{\hat{X}(\tilde{\lambda})}))((W_2(p_X,p_{\tilde{X}})\wedge P)-W_2(p_X,p_{\hat{X}(\tilde{\lambda})}))+\epsilon\nonumber\\
	&\leq 2W_2(p_{X},p_{\tilde{X}})((W_2(p_X,p_{\tilde{X}})\wedge P)-W_2(p_X,p_{\hat{X}(\tilde{\lambda})}))+\epsilon\nonumber\\
	&\leq 2\sigma_X((\sigma_X\wedge\sqrt{P})-(\sigma_X\wedge\sqrt{\tilde{P}}))+\epsilon,\label{eq:arbitrary}
\end{align}
where ($a$) is due to (\ref{eq:tight}) and (\ref{eq:bound}) in Lemma \ref{lem:DPtradeoff}. This proves
\begin{align}
	D(R,R_c,\tilde{P}|W^2_2)-D(R,R_c,P|W^2_2)\leq 2\sigma_X((\sigma_X\wedge\sqrt{P})-(\sigma_X\wedge\sqrt{\tilde{P}})),
\end{align}
which, together with the fact that the map $P\mapsto D(R,R_c,P|W^2_2)$ is monotonically decreasing, implies the desired continuity property.

\section{Proof of Theorem \ref{thm:connection}}\label{app:connection}



Note that for any $p_{\hat{X}|X}$ such that $I(X;\hat{X})\leq R$ and $H(\hat{X})\leq R+R_c$, the induced $p_{U\hat{X}|X}$ with $U:=\hat{X}$ satisfies (\ref{eq:constraint1})--(\ref{eq:constraint3}). This implies $D_e(R,R_c)\geq D(R,R_c,\infty)$. On the other hand, for any $p_{U\hat{X}|X}$ satisfying
(\ref{eq:constraint1})--(\ref{eq:constraint3}), it follows by the data processing inequality \cite[Theorem 2.8.1]{CT91} that $I(X;\hat{X})\leq R$. 	
Therefore, we must have $D(R,R_c,\infty)\geq D_e(R,\infty)$. This completes the proof of (\ref{eq:order}).

For the purpose of establishing the equivalence relationship (\ref{eq:imply}), it suffices to show that $D_e(R,R_c)>D_e(R,\infty)$ implies $D(R,R_c,\infty)>D_e(R,\infty)$ since the converse is implied by (\ref{eq:order}). 
To this end, we shall prove the contrapositive statement, namely, $D(R,R_c,\infty)\leq D_e(R,\infty)$ implies $D_e(R,R_c)\leq D_e(R,\infty)$. 	
Assume that the infimum in (\ref{eq:DRP}) is attained by some $p_{U^*\hat{X}^*|X}$. Let $\hat{U}^*:=\mathbb{E}[X|U^*]$. 
We have
\begin{align}
	D(R,R_c,\infty)&=\mathbb{E}[(X-\hat{X}^*)^2]\nonumber\\
	&\stackrel{(a)}{=}\mathbb{E}[(X-\hat{U}^*)^2]+\mathbb{E}[(\hat{X}^*-\hat{U}^*)^2],
\end{align}
where ($a$) holds because $\hat{U}^*=\mathbb{E}[X|U^*,\hat{X}^*]$ almost surely. Since	
\begin{align}
	I(X;\hat{U}^*)\leq I(X;U^*)\leq R,
\end{align}
it follows that $\mathbb{E}[(X-\hat{U}^*)^2]\geq D_e(R,\infty)$. Therefore, $D(R,R_c,\infty)\leq D_e(R,\infty)$ implies $\mathbb{E}[(X-\hat{U}^*)^2]=D_e(R,\infty)$ and $\mathbb{E}[(\hat{U}^*-\hat{X}^*)^2]=0$ (i.e., $\hat{U}^*=\hat{X}^*$ almost surely).
Note that
\begin{align}
	I(X;\hat{U}^*)\leq I(X;U^*)&\leq R
\end{align}
and
\begin{align}
	H(\hat{U}^*)=I(\hat{X}^*; \hat{U}^*)\leq I(\hat{X}^*; U^*)\leq R+R_c.
\end{align}
As a consequence, we have $\mathbb{E}[(X-\hat{U}^*)^2]\geq D_e(R,R_c)$. This proves $D_e(R,R_c)\leq D_e(R,\infty)$.

\section{Proof of Theorem \ref{thm:nontightness}}\label{app:nontightness}

\begin{lemma}
	For $p_X=\mathcal{N}(\mu_X,\sigma^2_X)$,
	\begin{align}
		D_e(R,R_c)>\underline{D}(R,R_c,\infty)\label{eq:Delower}
	\end{align}
	when $R\in(0,\infty)$ and $R_c\in[0,\infty)$, and 
	\begin{align}
		D_e(R,R_c)<\overline{D}(R,R_c,\infty)\label{eq:Deupper}
	\end{align}
	when $R_c\in[0,\infty)$ and $R\in(0,\chi(R_c))$, where $\chi(R_c)$ is a positive threshold that depends on $R_c$.
\end{lemma}
\begin{IEEEproof}
	Let $p_{\hat{X}^*|X}$ be some conditional distribution that attains the minimum in (\ref{eq:ECSQ}). Clearly, we must have $\mu_{\hat{X}^*}=\mu_X$. Note that
	\begin{align}
		R&\geq I(X;\hat{X}^*)\nonumber\\
		&=h(X)-h(X|\hat{X}^*)\nonumber\\
		&=h(X)-h(X-\hat{X}^*|\hat{X}^*)\nonumber\\
		&\stackrel{(a)}{\geq} h(X)-h(X-\hat{X}^*)\nonumber\\
		&\stackrel{(b)}{\geq}\frac{1}{2}\log\frac{\sigma^2_X}{D_e(R,R_c)}.
	\end{align}
	The inequalities ($a$) and ($b$) become equalities if and only if $X-\hat{X}^*$ is independent of $X^*$ and is distributed as $\mathcal{N}(0,D_e(R,R_c))$, which, together with the fact $p_X=\mathcal{N}(\mu_X,\sigma^2_X)$, 
	implies $p_{\hat{X}^*}=\mathcal{N}(\mu_X,\sigma^2_X-D_e(R,R_c))$.
	This is impossible since $H(\hat{X}^*)\leq R+R_c<\infty$ whereas the entropy of a Gaussian distribution with positive variance\footnote{Since $R>0$, it follows that $D_e(R,R_c)<\sigma^2_X$.} is infinite. Therefore, at least one of the inequalities ($a$) and ($b$) is strict, yielding
	\begin{align}
		R>\frac{1}{2}\log\frac{\sigma^2_X}{D_e(R,R_c)}.
	\end{align}
	Now one can readily prove (\ref{eq:Delower}) by invoking (\ref{eq:underlineD}).

	According to \cite[Theorems 9 and 12]{MN06}, 
	\begin{align}
		D_e(R,0)=\sigma^2_X(1-2R)+o(R),\label{eq:combine1}
	\end{align}
where $o(R)$ stands for a term that approaches zero more rapidly than $R$ as $R\rightarrow 0$.
	On the other hand, it can be deduced from (\ref{eq:overlineD}) that
	\begin{align}
		\overline{D}(R,R_c,\infty)=\sigma^2_X(1-2R+2Re^{-2R_c})+o(R).\label{eq:combine2}
	\end{align}
	Combining (\ref{eq:combine1}) and (\ref{eq:combine2}) then invoking the fact $D_e(R,R_c)\leq D_e(R,0)$ proves (\ref{eq:Deupper}).
\end{IEEEproof}

Clearly, (\ref{eq:Doverline}) is a direct consequence of (\ref{eq:order}) and (\ref{eq:Deupper}). Since 
\begin{align}
	\underline{D}(R,R_c,\infty)=D_e(R,\infty)
\end{align}
for $p_X=\mathcal{N}(\mu_X,\sigma^2_X)$, it is tempting to deduce (\ref{eq:Dunderline}) from  (\ref{eq:imply}) and (\ref{eq:Delower}). Unfortunately, (\ref{eq:imply}) relies on the assumption that the infimum in (\ref{eq:DRP}) can be attained, which has not been verified for Gaussian $p_X$. Nevertheless, we show below that the key idea underlying the proof of (\ref{eq:imply}) and (\ref{eq:Delower}), namely, violating (\ref{eq:Dunderline}) necessarily forces $\hat{U}^*$ to be Gaussian and coincide with $\hat{X}^*$, can be salvaged without resorting to the aforementioned assumption 
 by treating $(\hat{U}^*,\hat{X}^*)$  as a certain limit under weak convergence.

Assume that (\ref{eq:Dunderline}) does not hold, i.e.,  
\begin{align}
	D(R,R_c,\infty)=\sigma^2_Xe^{-2R}.
\end{align}
For any positive integer $k$, there exists $p_{U^{(k)}\hat{X}^{(k)}|X}$ satisfying
\begin{align}
	&I(X;U^{(k)})\leq R,\label{eq:con1}\\
	&I(\hat{X}^{(k)};U^{(k)})\leq R+R_c\label{eq:con2}
\end{align}
as well as the Markov chain constraint $X\leftrightarrow U^{(k)}\leftrightarrow \hat{X}^{(k)}$
such that
\begin{align}
	\mathbb{E}[(X-\hat{X}^{(k)})^2]\leq \sigma^2_Xe^{-2R}+\frac{1}{k}.\label{eq:distortionbound}
\end{align}
Let $\hat{U}^{(k)}:=\mathbb{E}[X|U^{(k)}]$ and $V^{(k)}:=X-\hat{U}^{(k)}$. Since $X\leftrightarrow U^{(k)}\leftrightarrow \hat{X}^{(k)}$ form a Markov chain, it follows that
\begin{align}
	\mathbb{E}[(X-\hat{X}^{(k)})^2]=\sigma^2_{V^{(k)}}+\mathbb{E}[(\hat{U}^{(k)}-\hat{X}^{(k)})^2],\label{eq:decomposition}
\end{align}
which, together with (\ref{eq:distortionbound}), implies
\begin{align}
	\sigma^2_{V^{(k)}}\leq\sigma^2_Xe^{-2R}+\frac{1}{k}.\label{eq:sigmaVupper}
\end{align}
Moreover, we have
\begin{align}
 h(V^{(k)}|\hat{U}^{(k)})	&\geq h(V^{(k)}|U^{(k)})\nonumber\\
 &=h(X|U^{(k)})\nonumber\\
	&=h(X)-I(X;U^{(k)})\nonumber\\
	&\geq h(X)-R\nonumber\\
	&=\frac{1}{2}\log(2\pi e\sigma^2_Xe^{-2R}),\label{eq:chVlower}
\end{align}
and consequently
\begin{align}
	h(V^{(k)})\geq\frac{1}{2}\log(2\pi e\sigma^2_Xe^{-2R}).\label{eq:hVlower}
\end{align}
 Combining (\ref{eq:sigmaVupper}) and (\ref{eq:hVlower}) gives
\begin{align}
	\phi_{KL}(p_{V^{(k)}}\|\mathcal{N}(0,\sigma^2_Xe^{-2R}))
	&=-h(V^{(k)})+\frac{1}{2}\log(2\pi\sigma^2_Xe^{-2R})+\frac{\sigma^2_{V^{(k)}}}{2\sigma^2_Xe^{-2R}}\nonumber\\
	&\leq\frac{1}{2k\sigma^2_X}e^{2R}.
\end{align}
Therefore, $p_{V^{(k)}}$ converges to $\mathcal{N}(0,\sigma^2_Xe^{-2R})$ in Kullback-Leibler divergence as $k\rightarrow\infty$. It can be shown that the sequence $\{p_{X\hat{U}^{(k)}V^{(k)}\hat{X}^{(k)}}\}_{k=1}^{\infty}$ is tight (cf. the proof of Theorem \ref{thm:property}). By Prokhorov's theorem \cite[Theorem 4]{GN14}, there exists a subsequence $\{p_{X\hat{U}^{(k_m)}V^{(k_m)}\hat{X}^{(k_m)}}\}_{m=1}^{\infty}$ converging weakly to some distribution $p_{X\hat{U}^*V^*\hat{X}^*}$. Clearly, we have $p_{V^*}=\mathcal{N}(0,\sigma^2_Xe^{-2R})$. Note that (\ref{eq:sigmaVupper}) implies
\begin{align}
	h(V^{(k)})\leq\frac{1}{2}\log(2\pi e(\sigma^2_Xe^{-2R}+\frac{1}{k})),
\end{align}
which, together with (\ref{eq:chVlower}), yields
\begin{align}
	I(\hat{U}^{(k)};V^{(k)})\leq\frac{1}{2}\log(1+\frac{1}{k\sigma^2_X}e^{2R}).
\end{align}
By the lower semicontinuity of mutual informaiton in the topology of weak convergence,
\begin{align}
	I(\hat{U}^*;V^*)\leq\liminf\limits_{m\rightarrow\infty}I(\hat{U}^{(k_m)};V^{(k_m)})=0.
\end{align}
Thus $\hat{U}^*$ and $V^*$ must be independent. Since $p_{\hat{U}^*+V^*}=p_X=\mathcal{N}(\mu_X,\sigma^2_X)$ and $p_{V^*}=\mathcal{N}(0,\sigma^2_Xe^{-2R})$, it follows that $p_{\hat{U}^*}=\mathcal{N}(\mu_X,\sigma^2_X(1-e^{-2R}))$. 

It remains to show that $\hat{U}^*=\hat{X}^*$ almost surely. In view of (\ref{eq:con1}),
the Shannon lower bound gives
\begin{align}
	\sigma^2_{V^{(k)}}\geq\sigma^2_Xe^{-2R},
\end{align}
which, together with (\ref{eq:distortionbound}) and (\ref{eq:decomposition}), further implies
\begin{align}
	\mathbb{E}[(\hat{U}^{(k)}-\hat{X}^{(k)})^2]\leq\frac{1}{k}.
\end{align}
As the map $(\hat{u},\hat{x})\mapsto (\hat{u}-\hat{x})^2$ is continuous and bounded from below, 
\begin{align}
	\mathbb{E}[(\hat{U}^*-\hat{X}^*)^2]\leq\liminf\limits_{m\rightarrow\infty}\mathbb{E}[(\hat{U}^{(k_m)}-\hat{X}^{(k_m)})^2]=0.
\end{align}
This leads to a contradiction with (\ref{eq:con2}) since
\begin{align}
	\liminf\limits_{k\rightarrow\infty}I(\hat{X}^{(k_m)};U^{(k_m)})&\stackrel{(a)}{\geq}\liminf\limits_{k\rightarrow\infty}I(\hat{X}^{(k_m)};\hat{U}^{(k_m)})\nonumber\\
	&\stackrel{(b)}{\geq} I(\hat{X}^*;\hat{U}^*)\nonumber\\
	&=\infty,
\end{align}
where ($a$) is due to the data processing inequality \cite[Theorem 2.8.1]{CT91}, and ($b$) is due to the lower semicontinuity of mutual informaiton in the topology of weak convergence.

The above proof can be simplified by circumventing the steps regarding the convergence of $p_{V^{(k)}}$ to $\mathcal{N}(0,\sigma^2_Xe^{-2R})$ in Kullback-Leibler divergence. Indeed, by
Cram\'{e}r's decomposition theorem, both  $\hat{U}^*$ and $V^*$ must be Gaussian if they are independent and their sum is Gaussian. Moreover, one can invoke the weak convergence argument to show that $\sigma^2_{\hat{U}^*}\leq\sigma^2_X(1-e^{-2R})$ and 
$\sigma^2_{V^*}\leq\sigma^2_Xe^{-2R}$. Since $\sigma^2_{\hat{U}^*}+\sigma^2_{V^*}=\sigma^2_X$, we must have $p_{\hat{U}^*}=\mathcal{N}(\mu_X,\sigma^2(1-e^{-2R}))$ and $p_{V^*}=\mathcal{N}(0,\sigma^2_Xe^{-2R})$. However, the original proof provides more information as convergence in Kullback-Leibler divergence is stronger than weak convergence.




\section{Proof of Corollary \ref{cor:weakperceptionKL}}\label{app:weakperceptionKL}

In light of Theorem \ref{thm:nontightness},
\begin{align}
	D(R,R_c,\infty|\phi_{KL})>\underline{D}(R,R_c,\infty|\phi_{KL})
\end{align}
when $R\in(0,\infty)$ and $R_c\in[0,\infty)$. This implies that
(\ref{eq:KLlower}) holds for sufficiently large $P$ 
since $P\mapsto D(R,R_c,P|\phi_{KL})$ is monotonically decreasing while $P\mapsto \underline{D}(R,R_c,P|\phi_{KL})$ is continuous at $P=\infty$.

In light of Theorem \ref{thm:nontightness},
\begin{align}
	D(R,R_c,\infty|\phi_{KL})<\overline{D}(R,R_c,\infty|\phi_{KL})
\end{align}
when $R_c\in[0,\infty)$ and $R\in(0,\chi(R_c))$. 
This implies that (\ref{eq:KLupper}) holds for sufficiently large $P$ since
$P\mapsto D(R,R_c,P|\phi_{KL})$ is continuous at $P=\infty$ by Theorem \ref{cor:KLcontinuity} and  $P\mapsto\overline{D}(R,R_c,P|\phi_{KL})$ is monotonically decreasing.

\section{Proof of Corollary \ref{cor:weakperceptionW2}}\label{app:weakperceptionW2}

In light of Theorem \ref{thm:nontightness},
\begin{align}
	D(R,R_c,\infty|W^2_2)>\underline{D}'(R,R_c,\infty|W^2_2)
\end{align}
when $R\in(0,\infty)$ and $R_c\in[0,\infty)$. This implies
that (\ref{eq:W2lower}) holds 
for $P$ above a postive threshold $\gamma'(R,R_c)$ strictly less than $P'(R,R_c)$ since $P\mapsto D(R,R_c,P|W^2_2)$ is monotonically decreasing while
$P\mapsto\underline{D}'(R,R_c,P|W^2_2)$ is continuous and remains constant over the interval $[P'(R,R_c),\infty]$.

In light of Theorem \ref{thm:nontightness},
\begin{align}
	D(R,R_c,\infty|W^2_2)<\overline{D}(R,R_c,\infty|W^2_2)
\end{align}
when $R_c\in[0,\infty)$ and $R\in(0,\chi(R_c))$. 
This implies that (\ref{eq:W2upper}) holds for $P$ above a postive threshold $\gamma(R,R_c)$ strictly less than $P(R,R_c)$ since
$P\mapsto D(R,R_c,P|W^2_2)$ is continuous by Theorem \ref{cor:W2continuity} and remains constant over the interval $[P(R,R_c),\infty]$ 
while  $P\mapsto\overline{D}(R,R_c,P|W^2_2)$ is monotonically decreasing.



\ifCLASSOPTIONcaptionsoff
  \newpage
\fi

\end{document}